\documentclass{article}

\usepackage[utf8]{inputenc} %

\usepackage{graphicx}

\usepackage[T1]{fontenc}

\IfFileExists{headers/config/showoverfull.config}{
	\overfullrule=1cm
}{
}

\usepackage{marginnote}

\usepackage[backgroundcolor=none,linecolor=red,textsize=footnotesize]{todonotes}

\usepackage{etoolbox}

\newbool{includeappendix}
\setbool{includeappendix}{true} %
\IfFileExists{headers/config/noappendix.config}{
	\setbool{includeappendix}{false}
}{}

\newif\ifincludeappendixx
\ifbool{includeappendix}{
	\includeappendixxtrue
}{
	\includeappendixxfalse
}

\usepackage{xr} %
\usepackage{filecontents}

\ifbool{includeappendix}{}{
	\input{appendix-labels-loader}

	\externaldocument{appendix-labels}
}

\definecolor{dark_green}{HTML}{137d13}
\definecolor{mid_green}{HTML}{30a830}

\definecolor{mid_blue}{HTML}{4265d9}
\definecolor{mid_purple}{HTML}{9d42d9}

\definecolor{light_gray}{HTML}{e6e6e6}
\definecolor{medium_gray}{HTML}{d4d4d4}
\definecolor{dark_gray}{HTML}{c7c7c7}

\colorlet{c_extraction}{mid_green!40}
\colorlet{c_iter}{mid_blue!40}
\colorlet{c_val}{mid_purple!40}

\definecolor{my-full-blue}{HTML}{1F77B4}

\definecolor{my-full-orange}{HTML}{FF7F0E}

\definecolor{my-full-green}{HTML}{2CA02C}

\definecolor{my-full-red}{HTML}{d62728}

\definecolor{my-full-purple}{HTML}{9467bd}

\definecolor{my-full-brown}{HTML}{8c564b}

\definecolor{my-full-pink}{HTML}{e377c2}

\definecolor{my-full-gray}{HTML}{7f7f7f}

\definecolor{my-full-olive}{HTML}{bcbd22}

\definecolor{my-full-cyan}{HTML}{17becf}

\colorlet{my-blue}{my-full-blue!30}
\colorlet{my-orange}{my-full-orange!30}
\colorlet{my-green}{my-full-green!30}
\colorlet{my-red}{my-full-red!30}
\colorlet{my-purple}{my-full-purple!30}
\colorlet{my-brown}{my-full-brown!30}
\colorlet{my-pink}{my-full-pink!30}
\colorlet{my-gray}{my-full-gray!30}
\colorlet{my-olive}{my-full-olive!30}
\colorlet{my-cyan}{my-full-cyan!30}

\usepackage{tikz}

\usetikzlibrary{arrows}
\usetikzlibrary{arrows.meta}
\usetikzlibrary{automata}
\usetikzlibrary{calc}
\usetikzlibrary{backgrounds}
\usetikzlibrary{decorations.markings}
\usetikzlibrary{decorations.pathmorphing}
\usetikzlibrary{decorations.pathreplacing}
\usetikzlibrary{fit}
\usetikzlibrary{patterns}
\usetikzlibrary{positioning}
\usetikzlibrary{shadows}
\usetikzlibrary{shapes}
\usetikzlibrary{shapes.geometric}
\usetikzlibrary{backgrounds}

\usepackage{microtype}

\usepackage{hyperref}
\usepackage{color,soul}
\usepackage{ bbold }
\usepackage{multirow}
\usepackage{caption,tabularx,booktabs,xltabular}
\usepackage{graphicx}
\usepackage{tabu}
\usepackage{array}
\usepackage{siunitx}
\usepackage{caption}
\usepackage{subcaption}
\usepackage{fontawesome}
\usepackage{stackengine}
\usepackage{svg}
\usepackage{mathtools}
\usepackage{xspace}
\usepackage{wrapfig}
\usepackage{makecell}
\usepackage{placeins}
\usepackage{float}
\usepackage{pifont}
\usepackage{svg}
\usepackage[most]{tcolorbox}
\usepackage{array}
\usepackage{dsfont}
\usepackage{enumitem}
\usepackage{calc}
\usepackage{tcolorbox}
\usepackage{multicol}

\usepackage{amsmath}
\usepackage{amssymb}
\usepackage{mathtools}
\usepackage{amsthm}
\usepackage{xspace}
\theoremstyle{plain}

\theoremstyle{definition}

\theoremstyle{remark}

\newcommand{\yes}{\checkmark\xspace}

\newcommand{\no}{\ding{55}}

\newcommand{\agent}{\textsc{SetupAgent}\xspace}
\newcommand{\swa}{\textsc{SWA}\xspace}
\newcommand{\swab}{\textsc{SWA}-Bench\xspace}
\newcommand{\sweeb}{\textsc{SWEE}-Bench\xspace}
\newcommand{\swee}{\textsc{SWEE}\xspace}
\newcommand{\sweb}{\textsc{SWE}-Bench\xspace}
\newcommand{\swe}{\textsc{SWE}\xspace}
\newcommand{\execagent}{\textsc{ExecutionAgent}\xspace}

\newcommand{\gptfom}{\textsc{GPT-4o-mini}\xspace}
\newcommand{\gptfo}{\textsc{GPT-4o}\xspace}
\newcommand{\haiku}{\textsc{Haiku-3.5}\xspace}

\newcommand{\code}[1]{\texttt{#1}\xspace}

\newcommand{\ftp}{\ensuremath{F \! \to \! P}\xspace}

\newcommand{\ptp}{\ensuremath{P \! \to \! P}\xspace}
\newcommand{\ptf}{\ensuremath{P \! \to \! F}\xspace}
\newcommand{\ftf}{\ensuremath{F \! \to \! F}\xspace}

\newlength\myheight
\newlength\mydepth
\settototalheight\myheight{Xygp}
\newcommand*\inlinegraphics[1]{%
  \settototalheight\myheight{Xygp}%
  \settodepth\mydepth{Xygp}%
  \raisebox{-\mydepth}{\includegraphics[height=\myheight]{#1}}%
}

\newcommand{\robot}{\inlinegraphics{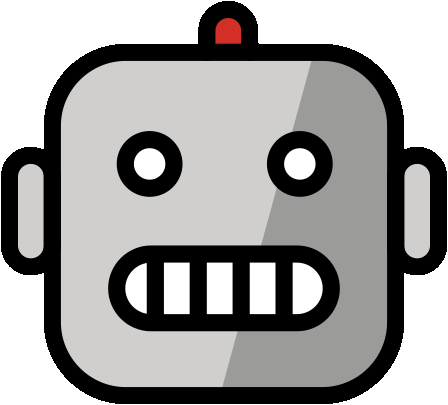}\xspace}
\newcommand{\comp}{\inlinegraphics{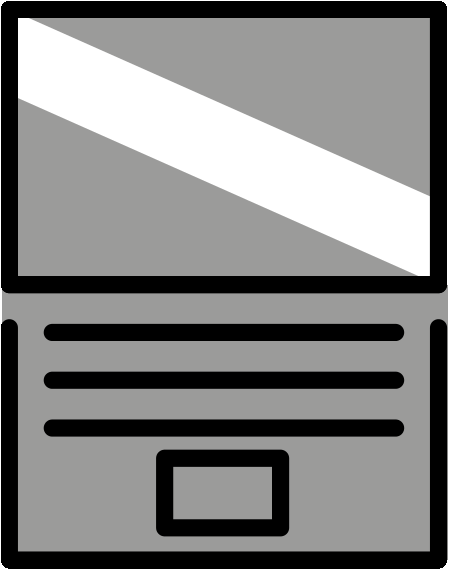}\xspace}

\newcommand{\prompt}[2]{\centering \begin{tcolorbox}[enhanced,width=0.9\textwidth,colback={light_gray},title={\textbf{#1}},colbacktitle=medium_gray,coltitle=black,frame hidden,boxrule=0pt]    
    \ttfamily \small #2
 \end{tcolorbox}}

 \newcommand{\markerb}[1]{\tikz[]{\node[fill, aspect=1, color=#1, inner sep=0pt, minimum size=2.1mm]{};}\xspace}
 \newcommand{\myexec}{\mathrm{exec}}

\usepackage[capitalize,noabbrev]{cleveref}

\makeatletter
\newcommand\theHALG@line{\thealgorithm.\arabic{ALG@line}}
\makeatother

\crefrangeformat{section}{\S#3#1#4\crefrangeconjunction\S#5#2#6}

\crefmultiformat{section}{\S#2#1#3}{\crefpairconjunction\S#2#1#3}{\crefmiddleconjunction\S#2#1#3}{\creflastconjunction\S#2#1#3}

\newcommand{\crefrangeconjunction}{--}

\crefname{listing}{Lst.}{listings}
\crefname{line}{Lin.}{Lin.}
\crefname{appendix}{App.}{App.}

\newcommand{\appref}[1]{%
	\ifbool{includeappendix}{\cref{#1}}{the appendix}%
}
\newcommand{\Appref}[1]{%
	\ifbool{includeappendix}{\cref{#1}}{The appendix}%
}

\usepackage[accepted]{icmlArxiv}

\icmltitlerunning{Automated Benchmark Generation for Repository-Level Coding Tasks}

\begin{document}

\twocolumn[
\icmltitle{Automated Benchmark Generation for Repository-Level Coding Tasks}\author{Author}

\icmlsetsymbol{equal}{*}
\begin{icmlauthorlist}
\icmlauthor{Konstantinos Vergopoulos}{lstar,equal}
\icmlauthor{Mark Niklas Müller}{lstar,equal}
\icmlauthor{Martin Vechev}{lstar,eth}
\end{icmlauthorlist}

\icmlaffiliation{eth}{Department of Computer Science, ETH Zurich}
\icmlaffiliation{lstar}{LogicStar AI}

\icmlcorrespondingauthor{Mark Niklas Müller}{mark@logicstar.ai}

\icmlkeywords{Machine Learning, ICML}

\vskip 0.3in
]

\printAffiliationsAndNotice{\icmlEqualContribution} %

\begin{abstract}
Code Agent development is an extremely active research area, where a reliable performance metric is critical for tracking progress and guiding new developments. This demand is underscored by the meteoric rise in popularity of \sweb. This benchmark challenges code agents to generate patches addressing GitHub issues given the full repository as context. The correctness of generated patches is then evaluated by executing a human-written test suite extracted from the repository after the issue's resolution.

However, constructing benchmarks like \sweb requires substantial manual effort to set up historically accurate execution environments for testing. Crucially, this severely limits the number of considered repositories, e.g., just 12 for \sweb. Considering so few repositories, selected for their popularity runs the risk of leading to a distributional mismatch, i.e., the measured performance may not be representative of real-world scenarios potentially misguiding development efforts. 

In this work, we address this challenge and introduce \agent, a fully automated system capable of historically accurate dependency setup, test execution, and result parsing. Using \agent, we generate two new datasets: (i) \sweeb an extended version of \sweb encompassing hundreds of repositories, and (ii) \swab a benchmark focusing on applications rather than libraries. Comparing these datasets to \sweb with respect to their characteristics and code agent performance, we find significant distributional differences, including lower issue description quality and detail level, higher fix complexity, and most importantly up to 40\% lower agent success rates.
\end{abstract}

\begin{figure*}[t]
    \centering
    \scalebox{0.93}{
        \input{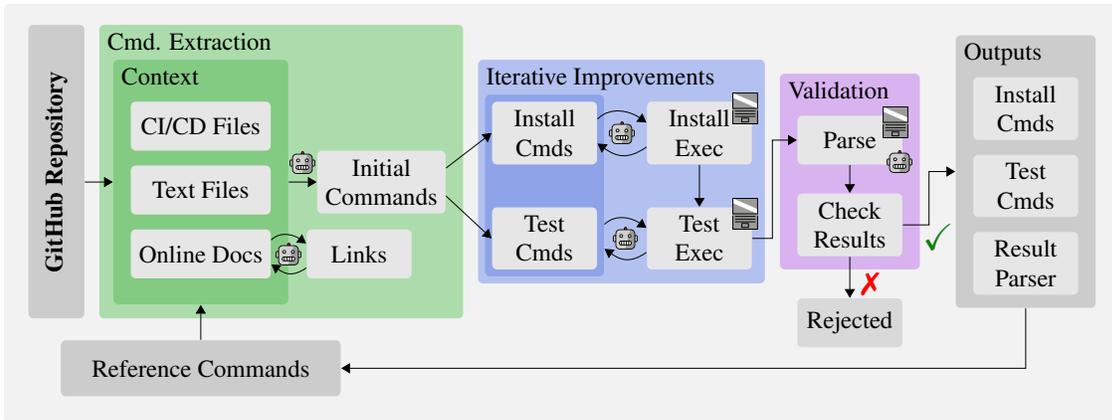}
    }
    \vspace{-3mm}
    \caption{Overview of \agent where a \robot-icon represents an LLM driven step and a \comp-icon represents execution feedback.}
    \label{fig:overview}
    \vspace{-4mm}
\end{figure*}

\newpage
\section{Introduction} \label{sec:introduction}
Code Agents are quickly becoming one of the most promising and actively researched applications of Large Language Models (LLMs); partly due to their potential to revolutionize the 700 billion dollar software industry \citep{statista2025}. To measure progress and more importantly steer further developments in this field, high-quality datasets and benchmarks are crucial. In particular, it is essential that they are representative of real-world use cases, sufficiently large to allow meaningful statistical analysis, and diverse and recent enough to avoid unintentional overfitting and contamination.

\vspace{-1mm}
\paragraph{Existing Benchmarks} However, function-level benchmarks like HumanEval \citep{HE2021}, popular for evaluating LLM's coding performance, are unrepresentative of real-world use, lack diversity, and are becoming saturated. To address these limitations, \sweb \citep{JimenezYWYPPN24} was proposed as the first repository-level coding benchmark based on real-world tasks, i.e., resolving GitHub issues. Yet, it still suffers from several limitations. (i) It is limited to few repositories, potentially leading to overfitting to these specific codebases. (ii) Its sole focus on libraries in contrast to applications raises generalizability questions. (iii) Its focus on popular repositories not only makes it less representative but also increases the chances of contamination with general codebase knowledge. (iv) Its static nature leads to most or even all instances being created before recent models' knowledge cutoff, allowing even the exact instances to be present in the training data.

\vspace{-1mm}
\paragraph{Creating Repository-Level Benchmarks} To address these challenges, we would like to create more diverse benchmarks and update them frequently with new tasks. However, while the GitHub Issues and Pull Requests (PRs), serving as task descriptions and reference solutions, respectively, for \sweb-like benchmarks can be scraped automatically, evaluating the correctness of a solution, requires the repository's test suite to be executed. This, in turn, requires setting up historically accurate execution environments, identifying the correct test commands, and parsing the results. Prior work addressed this problem either manually \citep{JimenezYWYPPN24} or by aggressively filtering out instances where default commands were unsuccessful \citep{jain2024r2e}. However, both approaches yield limited diversity and don't lend themselves to frequent updates.

\paragraph{This Work: \agent}
To address this challenge, we propose \agent, the first method to automate this setup process, enabling us to create repository-level code benchmarks fully automatically from a list of GitHub repositories.
\agent works in three key phases (illustrated in \cref{fig:overview}): (i) Command Extraction (green \markerb{c_extraction} in \cref{fig:overview}), (ii) Iterative Testing and Improvement (blue \markerb{c_iter}), and (iii) Validation (purple \markerb{c_val}). In the extraction phase, \agent analyzes relevant context, such as \code{README.md} files,  CI/CD configurations, and referenced web pages, to propose installation and testing commands. During the iterative improvement phase, \agent then executes these commands in a clean environment and leverages an LLM to systematically diagnose and resolve issues. Finally, in the validation phase,\agent ensures that the generated commands are reliable by verifying the correctness of the setup based on test results, only accepting configurations that meet a predefined success threshold.

\vspace{-1mm}
\paragraph{This Work: Generated Benchmarks} We demonstrate \agent's capability to generate coding benchmarks from a list of repositories by creating \swa- and \sweeb, each addressing specific shortcomings of \sweb. Both are designed to be representative of real-world use cases, consider many repositories leading to diverse benchmarks, and can be frequently updated without manual effort to avoid contamination and overfitting.
\swab focuses on software applications, containing 44 projects while \sweeb focuses on diversity and less popular projects containing 366 Python repositories. Comparing \swa- and \sweeb to \sweb, we find significant distributional differences, including lower repository age and popularity at issue creation, a larger focus on recent issues, and significantly more complex reference code fixes (2-4x more modified files and lines). Evaluating popular code agents on these datasets, we find significant performance differences for some models and statistically significant signs of contamination, highlighting the importance of evaluating on representative benchmarks.

\paragraph{Key Contributions} of this work are:
\vspace{-3mm}
\begin{itemize}
    \setlength\itemsep{0mm}
    \item We propose \agent, the first method for autonomously creating historically accurate execution environments.
    \item We leverage \agent to create two datasets for repository-level code generation \swa- and \sweeb, focusing on applications and diverse projects, respectively.
    \item We extensively analyze \swa- and \sweeb in terms of their characteristics and corresponding code agent performance.
\end{itemize}

\section{Related Work} \label{sec:related-work}
\vspace{-1mm}
\paragraph{Code Agents}
To fully leverage the potential of LLMs for code generation, they have been equipped with tools to interact with their environment without additional user input, e.g., by searching, viewing, and editing code, \citep{survey-agents}. These so-called code agents have shown great promise on complex tasks \citep{repairagent,opendevin,zhang2024autocoderover,yang2024sweagent,XiaDDZ24Agentless,Aider2024,RidnikKF24,openhands}. In this work, we evaluate some of the best-performing open-source agents.

\vspace{-1mm}
\paragraph{Code Generation Benchmarks}
With the success of LLMs in the domain of code generation, an increasing variety of function-level code generation benchmarks were proposed to assess their capabilities  \citep{HE2021,HendrycksBKMAGB21,MBPP2022,JainHGLYZWSSS24,HuangL0GLLLS0DC24}. However, not only were these increasingly saturated by state-of-the-art models but their focus on interview-style function-level coding challenges makes them also unrepresentative of the complexities of real-world codebases and software engineering tasks.

To address these limitations, a range of repository-level code-generation benchmarks have been proposed recently \citep{liu2023repobench,JainSZHSS24,JimenezYWYPPN24}. However, a repository-level context not only makes code generation more challenging but also dataset generation as it requires a historically accurate execution environment to be set up, the project's test suite to be run, and detailed results to be extracted. The required manual effort led to existing datasets focusing on a relatively small number of popular repositories. As a result, they are prone to overfitting, often lack diversity, and can easily contaminate the training data. 

\vspace{-1mm}
\paragraph{Automatic Dataset Generation}
These challenges could be addressed via automatic dataset generation, which has been successfully applied to function-level benchmarks by scraping tasks from coding challenge websites and doing varying levels of manual post-processing \citep{HendrycksBKMAGB21,JainHGLYZWSSS24,HuangL0GLLLS0DC24}. 

\citet{JimenezYWYPPN24} transfer these ideas to repository-level benchmarks, automatically scraping GitHub repositories, issues, and corresponding pull requests resolving these issues to create \sweb consisting of 12 repositories and 2294 instances. However, they still created the required execution environments and test commands manually.

\citet{JainSZHSS24} create \textsc{R2E}, a function-level synthesis benchmark with repository context by scraping GitHub repositories and masking out the function to be generated. They automated the setup by applying a default approach for projects with a \code{setup.py} or \code{pyproject.toml} file, automatically generating equivalence tests, and filtering out all instances where this approach fails. However, this approach aggressively filters projects with more complex installation procedures, not only introducing a selection bias but also yielding only 246 instances. %

In this work, we combine the more interesting repository-level tasks with a fully automated benchmark generation process, by introducing and leveraging \agent to automatically extract the installation and testing procedures for every task instance, allowing us to create larger and more diverse benchmarks efficiently.

\citet{BouzeniaP25}, concurrently proposed \execagent, a tool to automatically set up and test repositories. However, it is 60 times slower than \agent, does not support historical states, and does not extract results at test-level granularity. Even if the latter two shortcomings were addressed, it would remain infeasibly slow taking, e.g., over 4 months to generate \sweeb\footnote{Extrapolated from $\sim$150 repositories.}.

\vspace{-1mm}
\section{Autonomous Environment Setup} \label{sec:agent}
\vspace{-1mm}
In this Section, we first outline the requirements for a setup and testing agent to be used for benchmark generation and then describe the agent we develop for this purpose.

\subsection{Notation and Definitions}
\label{sec:notation_success}
We first introduce notation to describe repository-level coding tasks, adapted from \citet{mundler2024swt}.
Given a codebase $R$, we obtain $R \circ X$ by applying the code patch $X$.
We similarly denote the test suite $T$ with $T \circ S$ after applying the test patch $S$.
A single test $t \in T$ can either pass (P) or fail (F) when executed against the codebase $R$ in an execution environment $E$. We write: $\myexec_{E}(t, R) \in \{P, F\}$ and let the order $P > F$ hold.

A repository-level coding task can be written as the tuple $(R, T, I, E, S^*, X^*)$, where $R$ and $T$ are the original codebase and test suite, respectively, $I$ is the issue description, $E$ the execution environment, and $S^*$ and $X^*$ the reference test and code patch, respectively. By executing all tests $t_i \in T \circ S^*$ in the execution environment $E$, first against the original ($R$) and then the patched codebase ($R \circ X^*$), we obtain the reference test behavior $b^*_i = (\myexec_{E}(t_i, R) \to \myexec_{E}(t_i, R \circ X^*))$. We call $t_i$ with, e.g., $b^*_i = \ftp$ a fail-to-pass test as it fails before the reference fix is applied but passes afterward. 
We let the partial order $\ftp > \ftf$ and $\ptp > \ptf$ hold.

The task is now to generate a patch $X'$, given only $(R, T, I, E)$, such that the test behavior $b'_i = \myexec_{E}(t_i, R) \to \myexec_{E}(t_i, R \circ X')$ matches or improves on the reference result, i.e., $b'_i \geq b^*_i$ for all tests $t_i \in T \circ S^*$.

\subsection{Setup Agent Requirements}
A generic setup agent targeting individual, up-to-date repositories only has to satisfy one main requirement: \emph{Correctness} --  It must extract and run the installation and testing commands before parsing the test results.
However, benchmark generation, i.e., generating the execution environment $E$ given the remaining components of a coding task, imposes additional requirements: 
\emph{Historical Accuracy} -- Benchmark instances are based on specific, often outdated versions of a codebase $R$. The execution environment $E$ must thus use historically accurate dependency versions to reproduce the original issue faithfully and avoid version incompatibilities. 
\emph{Efficiency} -- To generate a dataset of many hundreds of instances, the setup agent must be efficient enough to keep total runtime reasonable (hours or at most few days). 
\emph{Granularity} -- Evaluating agent success requires test-level results to be parsed from the test suite output.

\vspace{-1mm}
\subsection{\agent}
\vspace{-1mm}
\paragraph{Overview} \agent works in three phases illustrated in \cref{fig:overview}: (1) Extraction (\markerb{c_extraction} in \cref{fig:overview}), (2) Iterative testing and imporvement (\markerb{c_iter}), and (3) Validation (\markerb{c_val}). In the first phase, \agent extracts a first version of the installation and testing commands from all relevant files, referenced webpages, and, if available, successful commands from similar versions of this repository. In the second phase, \agent iteratively executes first the installation and then testing commands, analyses the results and updates the commands. Finally, in the third phase, \agent validates the resulting commands by executing them, extracting the test results, and rejecting the proposed commands, if too few tests pass. Validated commands are then returned to the user and saved in a reference database to facilitate installations of different versions of the same repository.

\begin{figure}[h]
\centering
\begin{tcolorbox}[enhanced,width=0.98\linewidth,colback={light_gray},left=3pt,right=3pt,top=3pt,title={\ttfamily \scriptsize 
    \textbf{Input:}\\ Please extract all commands required to install <project\_name> in a clean environment and run its test suite from the context below.\\
    '{}'{}'\\
    \#\# README.md\\
    <file\_content>\\
    \#\# CONTRIBUTING.md\\
    <file\_content>\\
    '{}'{}'
    },colbacktitle=c_extraction,coltitle=black,frame hidden,boxrule=0pt]
    \ttfamily \scriptsize
    \textbf{LLM Response:}\\
    '{}'{}'bash\\
    apt-get install -y graphviz~~~\# installation\\
    pip install -r req.txt~~~~~~~~\# installation\\
    nox -e test~~~~~~~~~~~~~~~~~~~\# testing\\
    '{}'{}'\\
    \textbf{Installation Commands}:\\
    apt-get install -y graphviz\\
    uv pip install -r req.txt --exclude-newer 2022-04-11\\
    \textbf{Testing Commands}:\\
    nox -e test
\end{tcolorbox}
\vspace{-4mm}
\caption{Illustration of the extraction phase of \agent. Please see \cref{sec:app_prompts} for the full-length prompts.}
\vspace{-4mm}
\label{fig:extract_step}
\end{figure}

\paragraph{Extraction Phase} 
The goal of the Extraction Phase is to extract an initial version of the installation and testing commands.  We illustrate it in \cref{fig:extract_step}, using a significantly simplified prompt and shortened response for illustration purposes.
In the first step, \agent reviews all relevant files to identify the correct Python version. If available, it is directly parsed from the \code{setup.py} or \code{pyproject.toml} file, otherwise version constraints are extracted from different sources using an LLM and then resolved while taking issue creation time into account to ensure \emph{Historical Accuracy}.
Next, \agent uses a heuristic to identify possibly relevant context like \code{README.md} and \code{CONTRIBUTING.md} files, CI/CD configurations, and referenced webpages. It then reviews these sources and directly rejects those not containing information relevant to installation or testing, thus keeping its context size small and improving \emph{Efficiency}. Finally, \agent prompts an LLM to extract the installation and testing commands from the chosen sources.
To ensure \emph{Historical Accuracy}, \agent enforces the use of the \code{uv} environment manager for Python dependencies, configuring it to exclude dependency versions released after the issue creation.

\begin{figure}[h]
    \centering
    \begin{tcolorbox}[enhanced,width=0.98\linewidth,left=3pt,right=3pt,top=3pt,colback={light_gray},halign title=flush left, halign=flush left, title={\ttfamily \scriptsize 
        \textbf{Input:}\\ You are trying to install and test  <project\_name>  using the following commands.\\
        '{}'{}'bash\\
        <commands>\\
        '{}'{}'\\
        Modify the commands to resolve the following error:\\
        '{}'{}'\\
        \dots\\
        ModuleNotFoundError: No module named 'rustworkx'\\
        '{}'{}'
    },colbacktitle=c_iter,coltitle=black,frame hidden,boxrule=0pt]    
        \ttfamily \scriptsize
        \textbf{LLM Response:}\\
        We have to install rustworks first:\\
        '{}'{}'bash\\
        pip install rustworkx\\
        '{}'{}'\\
        \textbf{Updated Installation Commands}:\\
        apt-get install -y graphviz\\
        uv pip install -r req.txt \-\-exclude-newer 2022-04-11\\
        uv pip install rustworkx \-\-exclude-newer 2022-04-11
    \end{tcolorbox}
    \vspace{-4mm}
    \caption{Illustration of the iterative improvement phase of \agent, where the error message was obtained by executing the commands from the previous iteration.}
    \vspace{-4mm}
    \label{fig:iteration_step}
\end{figure}

\vspace{-1mm}
\paragraph{Iterative Improvement Phase} \agent now tries to install and test the project in a clean Docker container using the extracted commands. If any of the installation steps fail, i.e., exit with a non-zero exit code, or the test commands fail with an error, \agent selects the relevant sections of the error messages and then prompts an LLM to identify the issue and propose a modification of the installation commands to fix it. We illustrate this in \cref{fig:iteration_step}, where \agent adds a missing dependency.
Now \agent tries to resolve issues apparent in the test results, analyzing the logs to determine whether a failure is due to incorrect test commands, an incorrect installation, or a bug in the codebase. Depending on the result, \agent prompts an LLM to modify the installation or testing commands to fix the issue or passes the repo on to the validation phase. This iterative improvement is repeated until all errors are resolved or an iteration limit is reached. Using a moderate iteration limit of 4 steps, we achieve significantly improved \emph{Correctness} without sacrificing \emph{Efficiency}.

\begin{figure}[h]
    \centering
    \begin{tcolorbox}[enhanced,width=0.98\linewidth,left=3pt,right=3pt,top=3pt,colback={light_gray},title={\ttfamily \scriptsize 
        \textbf{Input:}\\ Please assess whether <project\_name> was installed and its test suite executed correctly given the resulting printout.\\
        Answer YES or NO.\\
        '{}'{}'\\
        ...\\
        ===== 2597 passed, 3 failed in 10.85s =====\\
        '{}'{}'
        },colbacktitle=c_val,coltitle=black,frame hidden,boxrule=0pt]    
        \ttfamily \scriptsize
        \textbf{LLM Response:}\\
        YES
    \end{tcolorbox}
    \vspace{-4mm}
    \caption{Illustration of the first step in the Validation phase.}
    \vspace{-4mm}
    \label{fig:val_step}
\end{figure}

\paragraph{Validation Phase} In the validation phase, \agent first queries an LLM to assess whether the installation and testing were successful, illustrated in \cref{fig:val_step}. If the LLM judges the installation to be successful, \agent updates the test framework's configuration to return test-level results, e.g., by adding \code{-rA} to a \code{pytest} command, thus ensuring \emph{Granularity}. It then selects the correct parser from a pre-defined set to extract test-level results and checks the number of passing and failing tests. We consider the installation to be successful if at least 95\% of tests pass.

\begin{figure*}[t]
    \centering
    \begin{subfigure}[t]{0.30\textwidth}
        \centering
        \includegraphics[width=1.05\textwidth]{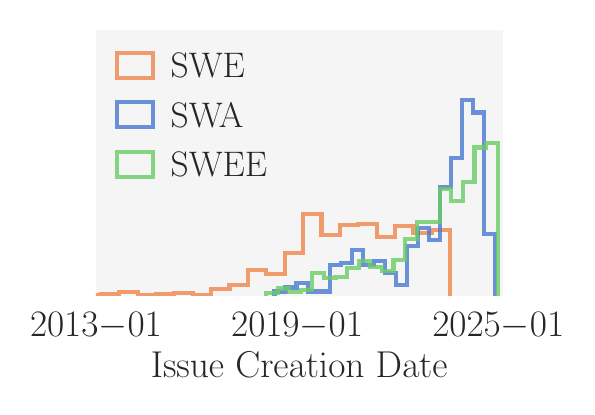}
    \end{subfigure}
    \hfill
    \begin{subfigure}[t]{0.30\textwidth}
        \centering
        \includegraphics[width=0.88\textwidth]{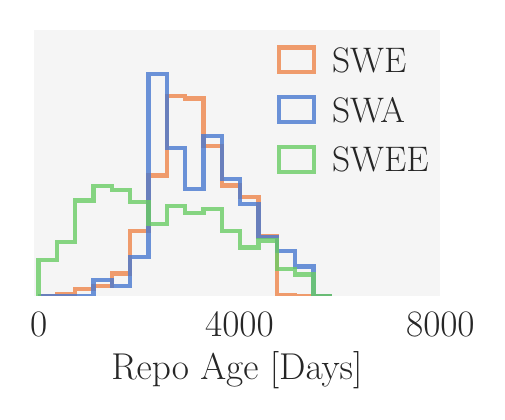}
    \end{subfigure}
    \hfill
    \begin{subfigure}[t]{0.30\textwidth}
        \centering
        \includegraphics[width=1.005\textwidth]{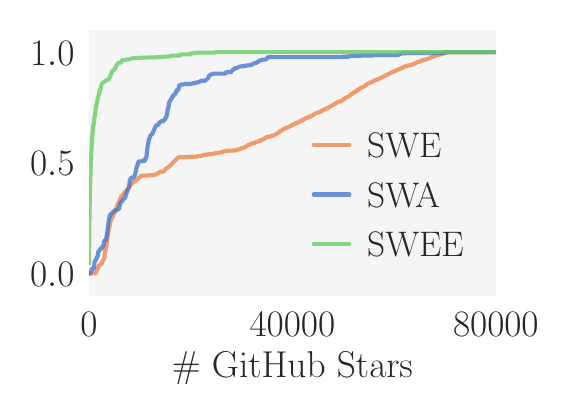}
    \end{subfigure}
    \vspace{-6mm}
    \caption{PDFs (left and middle) and CDF (right) of PR creation dates (left), repository age at PR creation time (middle), and number of GitHub stars (right) for \swa, \swee, and \sweb.}
    \vspace{-3mm}
    \label{fig:dataset_stats}
\end{figure*}

\section{Code Generation Benchmarks} \label{sec:dataset}
In this Section, we describe how we leverage \agent to create \swa- and \sweeb, two new benchmarks addressing specific limitations of \sweb. We compare these datasets with \sweb and provide insights into distributional differences.

\begin{figure}[h]
    \centering
    \includegraphics[width=0.8\linewidth]{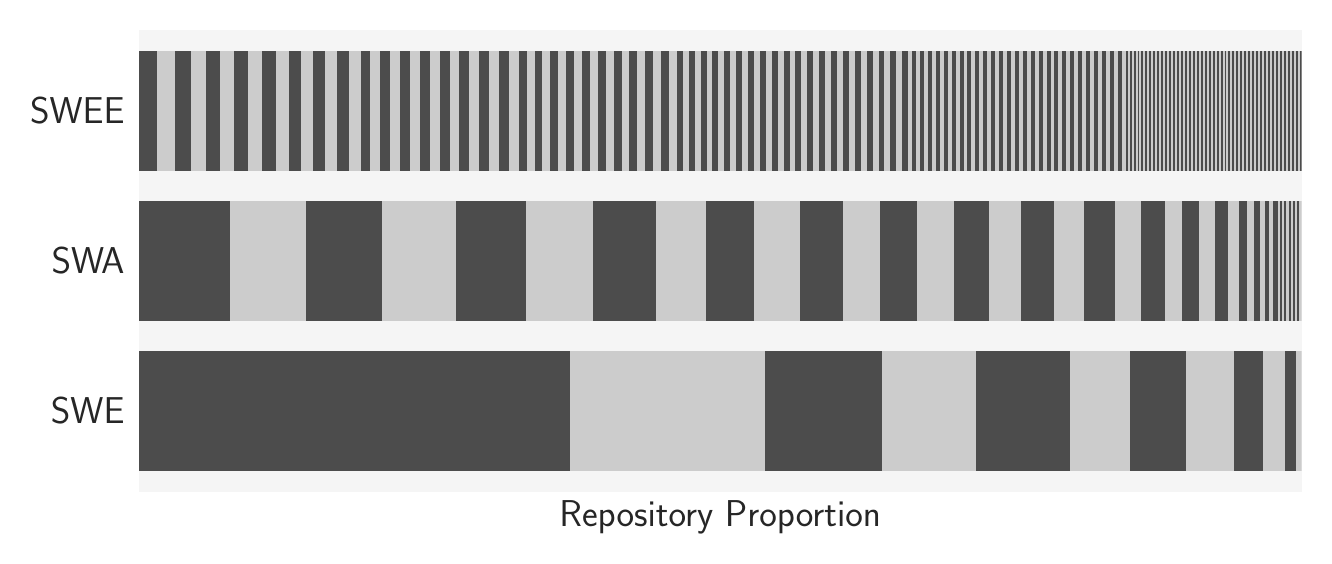}
    \vspace{-5mm}
    \caption{Comparison of the repository distribtuion of \swee-, \swa-, and \sweb across instances.}
    \label{fig:benchmark_composition}
\end{figure}

\paragraph{Automatically Generated Benchmarks} By creating execution environments automatically, we address two core limitations of manually generated repository-level benchmarks: (i) we can consider many more repositories without requiring infeasible manual labor, thus improving diversity and reducing the risk of overfitting and (ii) we can easily update benchmarks by creating new tasks from recent PRs and issues, thus ensuring that models are not contaminated with benchmark instances (see \cref{fig:dataset_stats,fig:benchmark_composition}).

\paragraph{\swab} Many practitioners using Code Agents develop software applications that suffer from different types of bugs compared to libraries due to architectural and structural differences. As \sweb only considers libraries, we design \swab to focus only on applications. 

\paragraph{\sweeb} We observe that more popular repositories tend to have higher-quality codebases and issue descriptions. This includes, e.g., a more consistent (file) structure and naming conventions, better documentation including detailed docstrings for most functions, and issue descriptions following a precise template (see \cref{fig:issue_stats,fig:fix_complexity}). As \sweb focuses on particularly popular Python repositories, the resulting tasks can be unrepresentative of real-world use. Therefore, we design \sweeb with a focus on diverse and less popular (median of 365 vs 16k stars) Python repositories (see \cref{fig:dataset_stats}).

\subsection{Dataset Creation}
\paragraph{Source Repositories}
For \swab, we combine a list of 468 popular Python applications \citep{Hashemi2024} with a list of 50 Python projects from \citet{BouzeniaKP24}, leading to a total of 475 candidate repositories after deduplication. For \sweeb, we consider the 8000 most downloaded PyPi projects at the time \citep{hugo_van_kemenade_2024_14252675} with between 100k and 1.5B monthly downloads and 0 to 25k stars, leading to good diversity while focusing on relevant projects.

\begin{table}[t]
    \centering
    \caption{\swee pipeline from projects to tasks. A PR is valid if it resolves an issue, modifies a test file, and is merged. An instance valid, if it has additionally at least one \ftp test.}
    \vspace{1mm}
    \label{tab:n_filters}
    \resizebox{0.8\linewidth}{!}{
    \begin{tabular}{llcc}
        \toprule
        Step & \# Repos & \# PRs \\
        \midrule
        Initial Projects & $8000$ &  \\
        $+$ GH Repo Found & $7057$ &  \\
        $+$ Preprocessing & $5097$ &  \\
        $+$ Permissive License & $3800$ &  \\
        $+$ Has valid PR & $2377$ & \\
        $+$ \agent succeeds & $514$ &  \\
        $+$ Get $n_{per\_repo}$ valid PRs & & $2115$ \\
        $+$ \agent succeeds & & $1513$ \\
        $+$ valid instance & & $885$ \\
        \bottomrule
    \end{tabular}
    }
    \vspace{-3mm}
\end{table}

\paragraph{Dataset Creation with \agent}
We combine the original PR filtering process from \citet{JimenezYWYPPN24} with our \agent as follows:
For every project, we first locate the corresponding repository, deduplicate the results, and filter out repositories that are not published under a permissive license. 
We then scrape issues and pull requests for each repository until we find the most recent PR that is merged, resolved an issue, and modified a test file. We call this a valid PR.
We then use \agent to set up an execution environment $E$ for the corresponding codebase $R$ (see \cref{sec:agent}).
For repositories where this succeeds, we scrape additional PRs until we have $n_{per\_repo}$ valid ones or, for \swee, reach a maximum of 500 PRs. 
We then use \agent to create the execution environment $E$ for each corresponding codebase $R$ in reverse chronological order per repository, populating \agent's reference commands database to speed up the setup process.
Finally, we split every PR into a reference code patch $X^*$ and test patch $S^*$. We execute the full test suite $T \circ S^*$ before and after the code patch is applied, i.e., on $R$ and $R\circ X^*$, respectively, to obtain the reference test behaviors $b_i^*$. We then filter out PRs, where test execution fails in one of these settings or which have no \ftp test, i.e. $\nexists\, t \in T \circ S^* : \myexec_{E}(t, R) \to \myexec_{E}(t, R \circ X^*) = \ftp$. The remaining PRs form the valid instances of the generated benchmark. We choose $n_{per\_repo}=50$ for \swab and $n_{per\_repo}=10$ for \sweeb to obtain the desired number of tasks and show the number of repositories and PRs this leads to in \cref{tab:n_filters_swa,tab:n_filters}, respectively.

\paragraph{Ease of Use}
To make benchmark generation and use as easy as possible, \agent only requires a list of repositories to generate a dataset in a format compatible with \sweb along with docker images with all dependencies installed. %

\subsection{Benchmark Characteristics}
\paragraph{Diversity} We compare the distribution of instances over repositories in \cref{fig:benchmark_composition} and observe that while instances in \swe are heavily concentrated in only a few repositories, with over 50\% of instances belonging to only two out of 12 total repositories, \swa- and \sweeb show much more diversity with 535 instances from 44 repositories and 885 from 366 repositories. See \cref{app:repo_list} for a full list.

\paragraph{Codebase Characteristics} We compare benchmarks with respect to codebase characteristics in \cref{tab:dataset_stats,fig:dataset_stats} and observe that \sweeb, compared to, \swa- and \sweb contains significantly older and more popular (\# GitHub stars) repositories and larger, more complex codebases (\# files and \# lines of code).

\begin{table}[ht]
    \centering
    \caption{Comparison of mean dataset characteristics.}
    \label{tab:dataset_stats}
    \resizebox{1\linewidth}{!}{
    \begin{tabular}{llrrr}
        \toprule
        & & \swa & \swee & \swe \\
        \midrule
        \multirow[1]{2}{*}{Codebase} 
        & \# Files & 899 & 77 & 1491\\
        & \# Lines & 112k & 14.8k & 321k\\
        \cmidrule(lr){1-1}
        \multirow[3]{3}{*}{Issue Descriptions}
        & \# Words & 240.2  & 125.1 & 181.3\\
        & \# Error Messages & 0.20 & 0.13 & 0.19\\
        & \# Code Blocks    & 1.53 & 1.19 & 1.06\\
        \cmidrule(lr){1-1}
        \multirow[3]{4}{*}{Tests}
        & \# \ptp & 564.2 & 226.6 & 120.1\\
        & \# \ftp & 38.8  & 38.1  & 13.5 \\
        & \# \ftf & 3.7   & 1.4   & 3.4\\
        & \# \ptf & 0.11  & 0.03  & 0.04\\
        \cmidrule(lr){1-1}
        \multirow[2.5]{4}{*}{Test Patches}
        & \# Edited Files & 1.89 & 2.05 & 1.52\\
        & \# Edited Lines & 74.8 & 91.5 & 39.2\\
        & \# Added Tests   & 9.10  & 23.78 & 6.37\\
        & \# Removed Tests & 16.77 & 2.49 & 0.54\\

        \cmidrule(lr){1-1}
        \multirow[2.5]{2}{*}{Fix Patches}
        & \# Edited Files & 3.26 & 3.26 & 1.66\\
        & \# Edited Lines & 104.3 & 169.9 & 41.0\\
        \bottomrule
    \end{tabular}
    }
\end{table}

\begin{figure}[t]
    \centering
    \hfill
    \begin{subfigure}[t]{0.32\linewidth}
        \centering
        \includegraphics[width=1.07\textwidth]{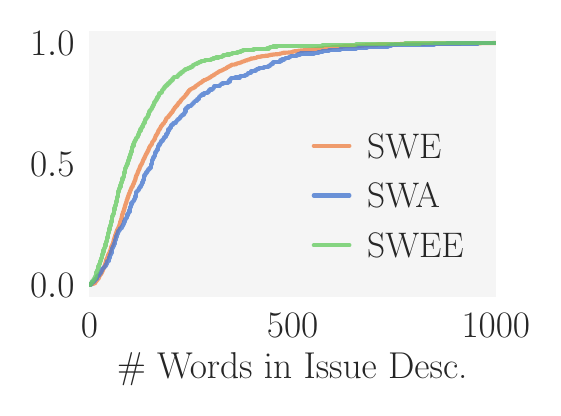}
    \end{subfigure}
    \hfill
    \begin{subfigure}[t]{0.32\linewidth}
        \centering
        \includegraphics[width=1.0\textwidth]{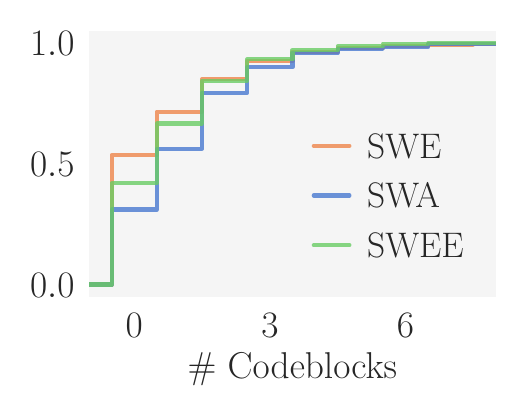}
    \end{subfigure}
    \hfill
    \begin{subfigure}[t]{0.32\linewidth}
        \centering
        \includegraphics[width=1.00\textwidth]{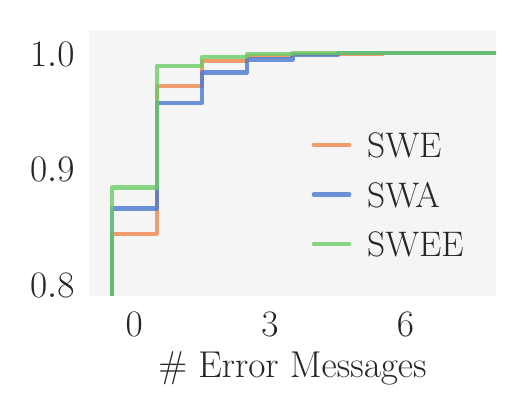}
    \end{subfigure}

    \begin{subfigure}[t]{0.32\linewidth}
        \centering
        \includegraphics[width=1.0\textwidth]{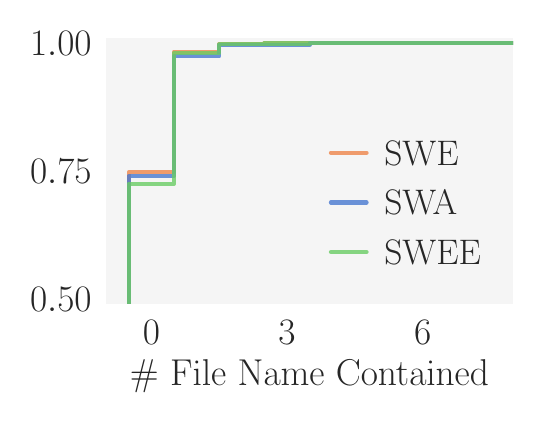}
    \end{subfigure}
    \hfill
    \begin{subfigure}[t]{0.32\linewidth}
        \centering
        \includegraphics[width=1.00\textwidth]{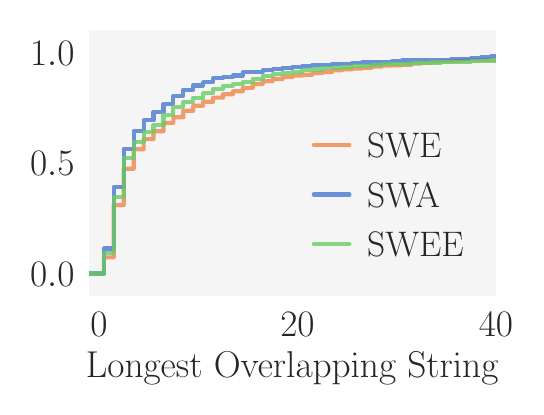}
    \end{subfigure}
    \hfill
    \begin{subfigure}[t]{0.32\linewidth}
        \centering
        \includegraphics[width=1.0\textwidth]{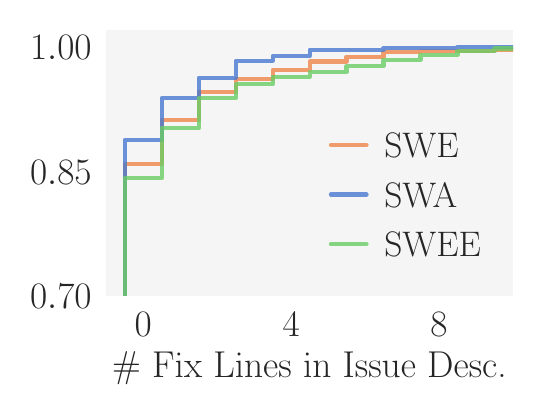}
    \end{subfigure}
    \vspace{-4mm}
    \caption{CDFs over issue description characteristics. Number of words (top left), number of code blocks (top middle), number of error messages (top right), number of filenames contained in the issue description and modified in the reference solution (bottom left), the overlap between the issue description and the reference solution in terms of longest string match (bottom middle) and complete lines (bottom right). A CDF further down and to the right indicates a higher value.}
    \label{fig:issue_stats}
\end{figure}

\paragraph{Issue Description Quality} To assess the issue description quality, we measure the number of words, error messages, and code blocks they contain as well as the overlap between the files mentioned there and modified in the reference fix and the overlap between the issue description and the reference solution itself. We show cumulative distribution functions (CDFs) of the aforementioned characteristics in \cref{fig:issue_stats}. We observe that while \swab has more detailed issue descriptions (longer, more code blocks, and more error messages), they do not seem to be of higher quality (less overlap with the reference solution and equal file mentions). Comparing \sweb and \sweeb, we observe longer issue descriptions and slightly more overlap with the reference solution in \sweb but otherwise similar characteristics.

\begin{figure}[t]
    \centering
    \hfill
    \begin{subfigure}[t]{0.32\linewidth}
        \centering
        \includegraphics[width=1.07\textwidth]{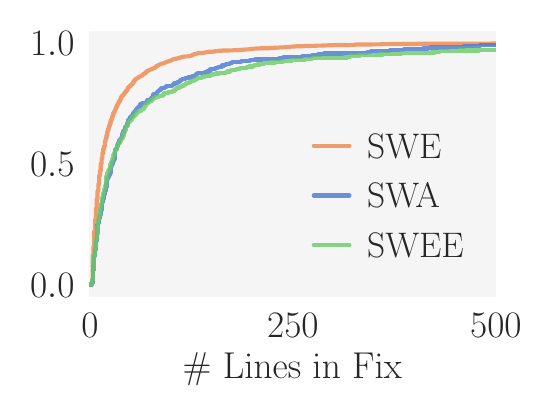}
    \end{subfigure}
    \hfill
    \begin{subfigure}[t]{0.32\linewidth}
        \centering
        \includegraphics[width=1.0\textwidth]{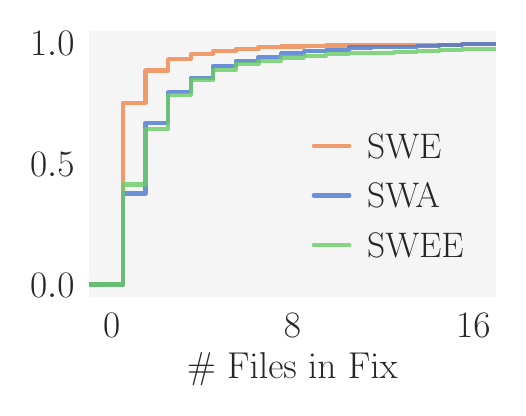}
    \end{subfigure}
    \hfill
    \begin{subfigure}[t]{0.32\linewidth}
        \centering
        \includegraphics[width=1.00\textwidth]{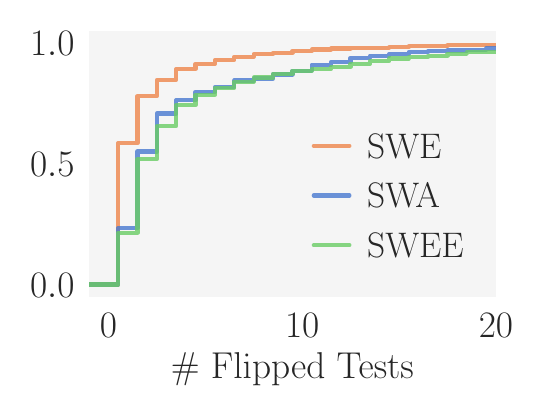}
    \end{subfigure}
    \vspace{-4mm}
    \caption{CDFs over fix-complexity characteristics. Number of edited lines (left), number of edited files (top middle), number affected tests, i.e., \ftp + \ptf (right). A CDF further down and to the right indicates higher characteristic values.}
    \label{fig:fix_complexity}
\end{figure}

\paragraph{Fix Complexity} To assess the complexity of required fixes, we measure the number of lines and files modified in the reference solution and the number of tests that flip from passing to failing (and vice versa). We show CDFs in \cref{fig:fix_complexity} and observe that while \swee- and \swab have similar distributions across all these metrics, \sweb fixes are significantly less complex by all metrics.

\section{Experimental Evaluation}\label{sec:experiments}
In this Section, we first evaluate the effectiveness of \agent for dataset creation and then analyze Code Agent performance across datasets.

\subsection{Experimental Setup}
\paragraph{Models}
We consider a range of models across sizes, cost points, and model providers. For exact versions, see \cref{tab:models} in \cref{sec:appendix-experiments}. Unless otherwise specified, we use \gptfom as the underlying model for all agents. For decoding, we use the default parameters for all Code Agents and greedy decoding for \agent.

\paragraph{Code Agents} 
We evaluate two state-of-the-art Code Agents from the top of the \sweb leaderboard\footnote{\href{https://www.swebench.com/\#verified}{swebench.com} accessed in November 2024} which most likely have been optimized for \sweb (OpenHands \citep{openhands} and AutoCodeRover-v2.0 \citep{zhang2024autocoderover}), and ZeroShot \citep{JimenezYWYPPN24} with oracle context (files modified in the ground truth fix) which prompts LLMs directly without any optimization for \sweb. %
We report the portion of resolved instances as accuracy (Acc.) for all Code Agents.

\paragraph{Code Execution} We run all code execution (both for \agent and all Code Agents) in separate Docker containers to improve reproducibility and security. For \agent, we use an Ubuntu 22.04 container as the base image and pre-install a range of common build dependencies but do not provide any Python dependencies.

\begin{table}[ht]
    \centering
    \caption{\agent success rates at extracting installation and test commands as well as parsing the resulting test output.}
    \vspace{1mm}
    \label{tab:agent_sucs}
    \resizebox{0.6\linewidth}{!}{
    \begin{tabular}{llr}
        \toprule
        & & Success \\
        \midrule
        \multirow{2}{*}{\swa} 
        & Repos  & $28.6\%$ \\
        & Instances & $58.5\%$ \\
        \cmidrule(lr){1-1}
        \multirow{2}{*}{\swee} 
        & Repos & $21.6\%$ \\
        & Instances & $71.5\%$\\
        \bottomrule
    \end{tabular}
    }
\end{table}

\subsection{Effectiveness of \agent}
We evaluate the effectiveness of \agent in creating \swa- and \sweeb by analyzing the frequency of fully successful environment and testing setups in \cref{tab:agent_sucs}. We observe \agent is able to extract historically correct execution environments for 20-30\% of repositories without reference commands and for 55-75\% of instances for these repositories.
Without reference commands, \agent takes 76 minutes to attempt to install all 154 repositories considered for \swa after deduplication and license checks and thus takes only about 30s on average per repository. When creating \sweeb, we deactivate the web browsing ability of \agent.

\begin{table}[t]
    \centering
    \caption{Ablation study on \agent, reporting the number of successfully extracted execution environments for \swab.}
    \vspace{1mm}
    \label{tab:ablation}
    \resizebox{0.7\linewidth}{!}{
    \begin{tabular}{lc}
        \toprule
        & \# Repositories \\
        \midrule
        \agent & 44 \\
        ~~only CI/CD Files & 33 \\
        ~~only Text Files & 15 \\
        ~~no Iterative Improvement & 11\\
        \bottomrule
    \end{tabular}
    }
    \vspace{-3mm}
\end{table}

\paragraph{Ablation} We evaluate the impact of \agent's components in an ablation study on \swab, reporting results in \cref{tab:ablation}. We observe that especially the use of CI/CD config files and the iterative improvement are crucial for \agent's success.

\paragraph{Failure Analaysis}To understand \agent's failure cases, we conduct a small case study, manually inspecting five failed instances from \swab, and observe the following: In all instances, errors in the build process cause the failure. For all but one instance, finding the installation instructions requires following two or more links on web pages. In all but two instances, the only described way to test the application requires running docker containers, which \agent does not support. In two instances, installation and/or testing requires the use of makefiles, referencing multiple substeps. Finally, in one instance \agent chooses the wrong requirement file and then begins to install missing testing dependencies.
We believe this points to exciting future work improving \agent's web-browsing capabilities and docker support.

\subsection{Agent Performance Across Datasets}
We conduct all below experiments on the full \swa and uniformly subsampled versions of \swee and \swe-Full of identical size (535 instances) due to cost constraints.

\begin{table}[t]
    \centering
    \caption{Issue resolution rates (accuracy) of various agents on \swa-, \swee-, and \sweb, all with \gptfom.}
    \label{tab:agent_perf}
    \vspace{1mm}
    \resizebox{0.7\linewidth}{!}{
    \begin{tabular}{lccc}
        \toprule
        & \swa & \swee & \swe \\
        \midrule
        Openhands         & $3.9\%$   & $4.4\%$     & $4.6\%$ \\
        AutoCodeRover v2  & $8.4\%$   & $8.9\%$     & $8.2\%$ \\ 
        ZeroShot(Oracle)  & $0.9\%$   & $2.2\%$     & $2.8\%$ \\
        \bottomrule
        \end{tabular}
    }
    \vspace{-3mm}
\end{table}

We report Code Agent performance in \cref{tab:agent_perf} and observe surprisingly small differences in performance between all three datasets when using \gptfom. %

To assess the interaction of agent performance and model selection, we evaluate AutoCodeRover v2 \citep{zhang2024autocoderover} across a range of LLMs, showing results in \cref{tab:model_perf}. Interestingly, we observe a large variance in the accuracy difference between \swe and \swa across models. While \gptfom performs similarly well on all benchmarks, \gptfo and \haiku perform much better on \swee- and \sweb. We show later that this may be due to contamination of \gptfo.

\begin{table}[ht]
    \centering
    \caption{Performance of AutocodeRover v2 \citep{zhang2024autocoderover} using different underlying LLMs.}
    \label{tab:model_perf}
    \vspace{1mm}
    \resizebox{0.7\linewidth}{!}{
    \begin{tabular}{lcccc}
    \toprule
    & \swa &  \swee & \swe \\
    \midrule
    \gptfom   & $8.4\%$  & $9.0\%$ & $8.2\%$ \\
    \gptfo    & $10.2\%$ &  $15.1\%$ & $16.6\%$ \\
    \haiku    & $10.8\%$  &  $12.9\%$ & 13.6\% \\ 
    \bottomrule
    \end{tabular}
    }
\end{table}

\subsection{Benchmark Analaysis}

In \cref{sec:dataset}, we observe interesting distributional differences between the instance characteristics of \swa-, \swee-, and \sweb. Now, we explore how these characteristics correlate with agent performance, reporting Spearman's rank correlation coefficients $\rho$ and p-values for AutoCodeRover v2 and \gptfo in \cref{tab:correlation}. We observe that only characteristics computed with knowledge of the solution have a statistically significant correlation with performance. In particular, the overlap of the issue with the reference code patch in terms of file names, and number of lines has a strong positive correlation with performance, while all fix complexity metrics have a strong negative correlation with performance.

\begin{table}[ht]
    \centering
    \caption{Spearman's rank correlation coefficients $\rho$ and p-value between accuracy and instance characteristics, separated by whether statistic can be computed without axes to ground truth. Statistically significant ($p<1\%$) correlations are highlighted in bold. Positive $\rho$ indicate that a larger characteristic value is associated with better performance.}
    \label{tab:correlation}
    \vspace{1mm}
    \resizebox{1\linewidth}{!}{
    \begin{tabular}{lcccccc}
    \toprule
    \multirow{2.5}{*}{Characteristic}& \multicolumn{2}{c}{\swa} & \multicolumn{2}{c}{\swee} & \multicolumn{2}{c}{\swe} \\
    \cmidrule(lr){2-3}
    \cmidrule(lr){4-5}
    \cmidrule(lr){6-7}
    & $\rho$ & p-value &  $\rho$ & p-value &  $\rho$ & p-value\\
    \midrule
    Repo Age                   & -0.06 & $2.0 \times10^{-1}$ & -0.02 & $5.8 \times 10^{-1}$  & -0.02 & $7.0 \times10^{-1}$\\
    \# GitHub Stars            & -0.03 & $4.8 \times10^{-1}$ & -0.02 & $7.2 \times 10^{-1}$ & ~0.07  & $1.3 \times10^{-1}$\\
    \# Words in Issue          & -0.06 & $1.8 \times10^{-1}$ & -0.00 & $9.6 \times 10^{-1}$ & ~0.01  & $7.7 \times10^{-1}$\\
    \# Code Blocks in Issue    & -0.00 & $9.7 \times10^{-1}$ & ~0.04 & $3.4 \times 10^{-1}$ & -0.06 & $ 1.7 \times10^{-1}$\\
    \# Error Messages in Issue & ~0.03  & $4.6 \times10^{-1}$ & ~0.09 & $3.7 \times 10^{-2}$ & -0.04   & $ 3.5 \times10^{-1}$\\
    \cmidrule(lr){1-1}
    \# Fix File Names in Issue & \textbf{~0.12}  & $4.5 \times10^{-3}$ & ~\textbf{0.19} & $1.2 \times 10^{-5}$ & \textbf{~0.18}  & $2.5 \times10^{-5}$\\
    Longest Fix Substring in Issue & -0.04 & $3.7 \times10^{-1}$& -0.11 & $1.1 \times 10^{-2}$ & ~0.04 & $3.1 \times10^{-1}$\\
    \# Fix Lines in Issue      & ~0.09  & $3.7 \times10^{-2}$ & ~0.06 & $1.7 \times 10^{-1}$  & \textbf{~0.17}   & $1.1 \times10^{-4}$\\
    \# Lines in Fix            & \textbf{-0.28} & ~$5.0 \times10^{-11}$ & \textbf{-0.40} & ~$1.3 \times 10^{-21}$  & \textbf{-0.28} & ~$6.2 \times10^{-11}$\\
    \# Files in Fix            & \textbf{-0.12} & $6.2 \times10^{-3}$ & \textbf{-0.26} & $1.6 \times 10^{-9}$ & \textbf{-0.16}  & $1.4 \times10^{-4}$\\
    \# Affected Tests          & \textbf{-0.18} & $3.3 \times10^{-5}$ & \textbf{-0.25} & $4.0 \times 10^{-9}$ & \textbf{-0.15}  & $7.2 \times10^{-4}$\\
    \bottomrule
    \end{tabular}
    }
\end{table}

\begin{table}[ht]
    \centering
    \caption{Accuracy of AutocodeRover v2 \citep{zhang2024autocoderover} on \swab instances split between those created before and after the model's knowledge cutoff (KC) and the p-value of the underlying resolution rate being the same or higher after the KC.}
    \label{tab:contamination}
    \vspace{1mm}
    \resizebox{1\linewidth}{!}{
    \begin{tabular}{llcccc}
    \toprule
    Dataset & Model & \# after KC & Acc before KC & Acc after KC & p-value\\
    \midrule
    \multirow{3}{*}{\swa}
    &\gptfom & 249 & 9.4\% & 7.2\% & 17.90\% \\
    &\gptfo & 249 & 12.2\% & 7.2\% & 2.65\% \\
    &\haiku & 44 & 11.0\% & 9.1\% & 34.83\% \\
    \cmidrule(lr){1-1}
    \multirow{3}{*}{\swee}
    &\gptfom & 230 & 8.2\% & 10.0\% & 76.50\%\\
    &\gptfo & 230 & 15.4\% & 15.2\% & 47.56\%\\
    &\haiku & 102 & 13.6\% & 9.8\% & 15.01\%\\
    \bottomrule
    \end{tabular}
    }
\end{table}

\paragraph{Data Contamination}
We analyze the accuracy (Acc) of AutoCodeRover v2 on \swa- and \sweeb, depending on whether a PR was created before or after a model's knowledge cutoff (KC), showing results in \cref{tab:contamination}. We report the (one-sided) p-value of observing these results under the null hypothesis that the success rate is not lower after the KC (computed using a t-test and normal approximation of the binomial distribution). We observe that on \swab all considered models have a lower success rate after the KC with the difference being statistically significant only for \gptfo. Interestingly, we observe no such signs on \sweeb which contains much less popular projects and is thus less prone to contamination. While all \swe instances are too old to conduct a similar analysis, we observe that the performance delta between \swe and \swa is correlated with the drop in accuracy over the KC on \swa.

\section{Conclusion} \label{sec:conclusion}
We introduced \agent, the first method for automated and historically accurate execution environment setup for Python codebases. \agent enables us to create repository-level code benchmarks fully automatically from a simple list of GitHub repositories. We demonstrated its effectiveness by creating two new benchmarks, \swa- and \sweeb, focusing on applications and diversity of codebases, respectively, and addressing several limitations of existing repository-level code benchmarks. In particular, their automated generation allows us to consider many more repositories, increasing diversity and reducing the risk of overfitting, and update the benchmarks over time, minimizing the risk of data contamination. 

We extensively analyzed \swa- and \sweeb, observing significant distributional differences compared to \sweb in fix-complexity characteristics that are strongly correlated with agent success. We further found statistically significant performance degradation for \swab instances created after the knowledge cutoff for one model. Together, these findings highlight the importance of evaluating on diverse, representative, and frequently updated benchmarks and thus the value of our automated benchmark generation approach. We believe \agent can facilitate this by enabling practitioners to quickly turn their specific target domain into a high-quality representative benchmark.

\message{^^JLASTBODYPAGE \thepage^^J}

\clearpage
\bibliography{references}

\begin{thebibliography}{26}
\providecommand{\natexlab}[1]{#1}
\providecommand{\url}[1]{\texttt{#1}}
\expandafter\ifx\csname urlstyle\endcsname\relax
  \providecommand{\doi}[1]{doi: #1}\else
  \providecommand{\doi}{doi: \begingroup \urlstyle{rm}\Url}\fi

\bibitem[Aider(2024)]{Aider2024}
Aider.
\newblock Aider is {SOTA} for both {SWE Bench} and {SWE Bench Lite}, Jun 2024.

\bibitem[Anthropic(2024)]{haiku2024}
Anthropic.
\newblock Model card addendum: Claude 3.5 haiku and upgraded claude 3.5 sonnet.
\newblock \url{https://assets.anthropic.com/m/1cd9d098ac3e6467/original/Claude-3-Model-Card-October-Addendum.pdf}, 2024.

\bibitem[Austin et~al.(2021)Austin, Odena, Nye, Bosma, Michalewski, Dohan, Jiang, Cai, Terry, Le, and Sutton]{MBPP2022}
Austin, J., Odena, A., Nye, M.~I., Bosma, M., Michalewski, H., Dohan, D., Jiang, E., Cai, C.~J., Terry, M., Le, Q.~V., and Sutton, C.
\newblock Program synthesis with large language models.
\newblock \emph{CoRR}, abs/2108.07732, 2021.
\newblock URL \url{https://arxiv.org/abs/2108.07732}.

\bibitem[Bouzenia \& Pradel(2024)Bouzenia and Pradel]{BouzeniaP25}
Bouzenia, I. and Pradel, M.
\newblock You name it, {I} run it: An {LLM} agent to execute tests of arbitrary projects.
\newblock \emph{CoRR}, abs/2412.10133, 2024.
\newblock \doi{10.48550/ARXIV.2412.10133}.
\newblock URL \url{https://doi.org/10.48550/arXiv.2412.10133}.

\bibitem[Bouzenia et~al.(2024{\natexlab{a}})Bouzenia, Devanbu, and Pradel]{repairagent}
Bouzenia, I., Devanbu, P.~T., and Pradel, M.
\newblock Repairagent: An autonomous, llm-based agent for program repair.
\newblock \emph{CoRR}, 2024{\natexlab{a}}.

\bibitem[Bouzenia et~al.(2024{\natexlab{b}})Bouzenia, Krishan, and Pradel]{BouzeniaKP24}
Bouzenia, I., Krishan, B.~P., and Pradel, M.
\newblock Dypybench: {A} benchmark of executable python software.
\newblock \emph{Proc. {ACM} Softw. Eng.}, 1\penalty0 ({FSE}):\penalty0 338--358, 2024{\natexlab{b}}.
\newblock \doi{10.1145/3643742}.
\newblock URL \url{https://doi.org/10.1145/3643742}.

\bibitem[Chen et~al.(2021)Chen, Tworek, Jun, Yuan, de~Oliveira~Pinto, Kaplan, Edwards, Burda, Joseph, Brockman, Ray, Puri, Krueger, Petrov, Khlaaf, Sastry, Mishkin, Chan, Gray, Ryder, Pavlov, Power, Kaiser, Bavarian, Winter, Tillet, Such, Cummings, Plappert, Chantzis, Barnes, Herbert{-}Voss, Guss, Nichol, Paino, Tezak, Tang, Babuschkin, Balaji, Jain, Saunders, Hesse, Carr, Leike, Achiam, Misra, Morikawa, Radford, Knight, Brundage, Murati, Mayer, Welinder, McGrew, Amodei, McCandlish, Sutskever, and Zaremba]{HE2021}
Chen, M., Tworek, J., Jun, H., Yuan, Q., de~Oliveira~Pinto, H.~P., Kaplan, J., Edwards, H., Burda, Y., Joseph, N., Brockman, G., Ray, A., Puri, R., Krueger, G., Petrov, M., Khlaaf, H., Sastry, G., Mishkin, P., Chan, B., Gray, S., Ryder, N., Pavlov, M., Power, A., Kaiser, L., Bavarian, M., Winter, C., Tillet, P., Such, F.~P., Cummings, D., Plappert, M., Chantzis, F., Barnes, E., Herbert{-}Voss, A., Guss, W.~H., Nichol, A., Paino, A., Tezak, N., Tang, J., Babuschkin, I., Balaji, S., Jain, S., Saunders, W., Hesse, C., Carr, A.~N., Leike, J., Achiam, J., Misra, V., Morikawa, E., Radford, A., Knight, M., Brundage, M., Murati, M., Mayer, K., Welinder, P., McGrew, B., Amodei, D., McCandlish, S., Sutskever, I., and Zaremba, W.
\newblock Evaluating large language models trained on code.
\newblock \emph{CoRR}, abs/2107.03374, 2021.
\newblock URL \url{https://arxiv.org/abs/2107.03374}.

\bibitem[Hashemi(2024)]{Hashemi2024}
Hashemi, M.
\newblock Awesome python applications.
\newblock \url{https://github.com/mahmoud/awesome-python-applications}, 2024.

\bibitem[Hendrycks et~al.(2021)Hendrycks, Basart, Kadavath, Mazeika, Arora, Guo, Burns, Puranik, He, Song, and Steinhardt]{HendrycksBKMAGB21}
Hendrycks, D., Basart, S., Kadavath, S., Mazeika, M., Arora, A., Guo, E., Burns, C., Puranik, S., He, H., Song, D., and Steinhardt, J.
\newblock Measuring coding challenge competence with {APPS}.
\newblock In Vanschoren, J. and Yeung, S. (eds.), \emph{Proceedings of the Neural Information Processing Systems Track on Datasets and Benchmarks 1, NeurIPS Datasets and Benchmarks 2021, December 2021, virtual}, 2021.
\newblock URL \url{https://datasets-benchmarks-proceedings.neurips.cc/paper/2021/hash/c24cd76e1ce41366a4bbe8a49b02a028-Abstract-round2.html}.

\bibitem[Huang et~al.(2024)Huang, Lin, Liu, Gong, Lu, Lei, Liang, Shen, Lin, Duan, and Chen]{HuangL0GLLLS0DC24}
Huang, Y., Lin, Z., Liu, X., Gong, Y., Lu, S., Lei, F., Liang, Y., Shen, Y., Lin, C., Duan, N., and Chen, W.
\newblock Competition-level problems are effective {LLM} evaluators.
\newblock In Ku, L., Martins, A., and Srikumar, V. (eds.), \emph{Findings of the Association for Computational Linguistics, {ACL} 2024, Bangkok, Thailand and virtual meeting, August 11-16, 2024}, pp.\  13526--13544. Association for Computational Linguistics, 2024.
\newblock \doi{10.18653/V1/2024.FINDINGS-ACL.803}.
\newblock URL \url{https://doi.org/10.18653/v1/2024.findings-acl.803}.

\bibitem[Jain et~al.(2024{\natexlab{a}})Jain, Han, Gu, Li, Yan, Zhang, Wang, Solar{-}Lezama, Sen, and Stoica]{JainHGLYZWSSS24}
Jain, N., Han, K., Gu, A., Li, W., Yan, F., Zhang, T., Wang, S., Solar{-}Lezama, A., Sen, K., and Stoica, I.
\newblock Livecodebench: Holistic and contamination free evaluation of large language models for code.
\newblock \emph{CoRR}, abs/2403.07974, 2024{\natexlab{a}}.
\newblock \doi{10.48550/ARXIV.2403.07974}.
\newblock URL \url{https://doi.org/10.48550/arXiv.2403.07974}.

\bibitem[Jain et~al.(2024{\natexlab{b}})Jain, Shetty, Zhang, Han, Sen, and Stoica]{JainSZHSS24}
Jain, N., Shetty, M., Zhang, T., Han, K., Sen, K., and Stoica, I.
\newblock {R2E:} turning any github repository into a programming agent environment.
\newblock In \emph{Forty-first International Conference on Machine Learning, {ICML} 2024, Vienna, Austria, July 21-27, 2024}. OpenReview.net, 2024{\natexlab{b}}.
\newblock URL \url{https://openreview.net/forum?id=kXHgEYFyf3}.

\bibitem[Jain et~al.(2024{\natexlab{c}})Jain, Shetty, Zhang, Han, Sen, and Stoica]{jain2024r2e}
Jain, N., Shetty, M., Zhang, T., Han, K., Sen, K., and Stoica, I.
\newblock R2e: Turning any github repository into a programming agent test environment.
\newblock In \emph{ICLR 2024}, 2024{\natexlab{c}}.

\bibitem[Jimenez et~al.(2024)Jimenez, Yang, Wettig, Yao, Pei, Press, and Narasimhan]{JimenezYWYPPN24}
Jimenez, C.~E., Yang, J., Wettig, A., Yao, S., Pei, K., Press, O., and Narasimhan, K.~R.
\newblock Swe-bench: Can language models resolve real-world github issues?
\newblock In \emph{The Twelfth International Conference on Learning Representations, {ICLR} 2024, Vienna, Austria, May 7-11, 2024}. OpenReview.net, 2024.
\newblock URL \url{https://openreview.net/forum?id=VTF8yNQM66}.

\bibitem[Liu et~al.(2023)Liu, Xu, and McAuley]{liu2023repobench}
Liu, T., Xu, C., and McAuley, J.~J.
\newblock Repobench: Benchmarking repository-level code auto-completion systems.
\newblock \emph{CoRR}, abs/2306.03091, 2023.
\newblock \doi{10.48550/ARXIV.2306.03091}.

\bibitem[M{\"u}ndler et~al.(2024)M{\"u}ndler, Mueller, He, and Vechev]{mundler2024swt}
M{\"u}ndler, N., Mueller, M.~N., He, J., and Vechev, M.
\newblock Swt-bench: Testing and validating real-world bug-fixes with code agents.
\newblock In \emph{The Thirty-eighth Annual Conference on Neural Information Processing Systems}, 2024.

\bibitem[OpenAI(2025)]{openai2025}
OpenAI.
\newblock Openai model docs.
\newblock \url{https://platform.openai.com/docs/models/gpt-4o }, 2025.

\bibitem[OpenDevin(2024)]{opendevin}
OpenDevin.
\newblock Opendevin: Code less, make more, 2024.

\bibitem[Ridnik et~al.(2024)Ridnik, Kredo, and Friedman]{RidnikKF24}
Ridnik, T., Kredo, D., and Friedman, I.
\newblock Code generation with alphacodium: From prompt engineering to flow engineering.
\newblock \emph{CoRR}, abs/2401.08500, 2024.
\newblock \doi{10.48550/ARXIV.2401.08500}.
\newblock URL \url{https://doi.org/10.48550/arXiv.2401.08500}.

\bibitem[Statista(2025)]{statista2025}
Statista.
\newblock Statista market insights.
\newblock \url{https://www.statista.com/outlook/tmo/software/worldwide}, 2025.

\bibitem[van Kemenade et~al.(2024)van Kemenade, Paterson, Thoma, Si, and Dollenstein]{hugo_van_kemenade_2024_14252675}
van Kemenade, H., Paterson, C., Thoma, M., Si, R., and Dollenstein, Z.
\newblock hugovk/top-pypi-packages: Release 2024.12, December 2024.
\newblock URL \url{https://doi.org/10.5281/zenodo.14252675}.

\bibitem[Wang et~al.(2024{\natexlab{a}})Wang, Ma, Feng, Zhang, Yang, Zhang, Chen, Tang, Chen, Lin, Zhao, Wei, and Wen]{survey-agents}
Wang, L., Ma, C., Feng, X., Zhang, Z., Yang, H., Zhang, J., Chen, Z., Tang, J., Chen, X., Lin, Y., Zhao, W.~X., Wei, Z., and Wen, J.
\newblock A survey on large language model based autonomous agents.
\newblock \emph{Frontiers Comput. Sci.}, 2024{\natexlab{a}}.

\bibitem[Wang et~al.(2024{\natexlab{b}})Wang, Li, Song, Xu, Tang, Zhuge, Pan, Song, Li, Singh, Tran, Li, Ma, Zheng, Qian, Shao, Muennighoff, Zhang, Hui, Lin, Brennan, Peng, Ji, and Neubig]{openhands}
Wang, X., Li, B., Song, Y., Xu, F.~F., Tang, X., Zhuge, M., Pan, J., Song, Y., Li, B., Singh, J., Tran, H.~H., Li, F., Ma, R., Zheng, M., Qian, B., Shao, Y., Muennighoff, N., Zhang, Y., Hui, B., Lin, J., Brennan, R., Peng, H., Ji, H., and Neubig, G.
\newblock {OpenHands: An Open Platform for AI Software Developers as Generalist Agents}, 2024{\natexlab{b}}.
\newblock URL \url{https://arxiv.org/abs/2407.16741}.

\bibitem[Xia et~al.(2024)Xia, Deng, Dunn, and Zhang]{XiaDDZ24Agentless}
Xia, C.~S., Deng, Y., Dunn, S., and Zhang, L.
\newblock Agentless: Demystifying llm-based software engineering agents.
\newblock \emph{CoRR}, abs/2407.01489, 2024.
\newblock \doi{10.48550/ARXIV.2407.01489}.
\newblock URL \url{https://doi.org/10.48550/arXiv.2407.01489}.

\bibitem[Yang et~al.(2024)Yang, Jimenez, Wettig, Lieret, Yao, Narasimhan, and Press]{yang2024sweagent}
Yang, J., Jimenez, C.~E., Wettig, A., Lieret, K., Yao, S., Narasimhan, K., and Press, O.
\newblock {SWE}-agent: Agent {Computer} {Interfaces} {Enable} {Software} {Engineering} {Language} {Models}, 2024.

\bibitem[Zhang et~al.(2024)Zhang, Ruan, Fan, and Roychoudhury]{zhang2024autocoderover}
Zhang, Y., Ruan, H., Fan, Z., and Roychoudhury, A.
\newblock Autocoderover: Autonomous program improvement.
\newblock \emph{CoRR}, abs/2404.05427, 2024.
\newblock \doi{10.48550/ARXIV.2404.05427}.

\end{thebibliography}
\bibliographystyle{icml2025}

\message{^^JLASTREFERENCESPAGE \thepage^^J}

\ifincludeappendixx
	\newpage
	\appendix
	\onecolumn
	\section{Appendix: Experiments} \label{sec:appendix-experiments}
Below, we provide the exact model versions we used in \cref{tab:models}.
\begin{table}[ht]
    \centering
    \caption{LLM Details inlcuding Knowledge Cutoff (KC)}
    \vspace{1mm}
    \label{tab:models}
    \resizebox{0.7\linewidth}{!}{
    \begin{tabular}{llccc}
        \toprule
        Model Name & Model ID & API Provider & KC & Reference\\
        \midrule
        \gptfo & \code{gpt-4o-2024-08-06} & OpenAI & Oct 2023 & \citet{openai2025}\\
        \gptfom & \code{gpt-4o-mini-2024-07-18} & OpenAI & Oct 2023 & \citet{openai2025}\\
        \haiku & \code{claude-3-5-haiku-20241022} & Anthropic & Jul 2024 & \citet{haiku2024} \\
        \bottomrule
    \end{tabular}
    }
\end{table}

\begin{table}[t]
    \centering
    \caption{\swa pipeline from projects to tasks. A PR is valid if it resolves an issue, modifies a test file, and is merged. An instance valid, if it has additionally at least one \ftp test.}
    \vspace{1mm}
    \label{tab:n_filters_swa}
        \begin{tabular}{llcc}
            \toprule
            Step & \# Repos & \# PRs \\
            \midrule
            Initial Projects & $475$ &  \\
            $+$ GH Repo Found & $440$ &  \\
            $+$ Preprocessing & $427$ &  \\
            $+$ Permissive License & $227$ &  \\
            $+$ Has valid PR & $154$ & \\
            $+$ \agent{} succeeds & $44$ &  \\
            $+$ Get up to 50 valid PRs & & $1527$ \\
            $+$ \agent{} succeeds & & $893$ \\
            $+$ valid instance & & $535$ \\
            \bottomrule
        \end{tabular}
    \vspace{-3mm}
\end{table}

\section{Appendix: Prompts} \label{sec:app_prompts}
In this Section, we provide the full-length prompts used by \agent.
\begin{figure}[H]
    \centering
    \prompt{Prompt to suggest relevant files}{
        You are a senior developer contributing to the www.github.com/<repo\_id> project by solving issues. You have created a Docker environment with Ubuntu, and now you want to install the repository in development mode (meant for active development and testing) and run the tests.
        The first step is to locate the installation instructions and the test commands. I will provide you a list of filenames or file paths (e.g., README.md, contributing.md), which typically include instructions for installation and testing.
        The files can be either filenames (e.g., README.md) or file paths (e.g., docs/maintaining/installing/install-from-source.rst).
        From the provided list of filenames or file paths your task is:
        1. Identify those likely related to installation or testing based on their names.
        2. Exclude those that are clearly irrelevant.
        3. If unsure, include the file/path in your response.
        4. Return only the files/paths from the given list, exactly as they appear, without modifying their names or structure
        5. If a full path is given, return the full path, not just the filename.
        6. Use the following format for your response
        <ANSWER>: file 1, \ldots file n, filepath 1, \ldots filepath k\\
        <REASONING>: <YOUR REASONING>\\
        Example input:\\
        `{}`{}`\\
        readme.md, contributing.md, contributors.md, docs/maintaining/installing/install-from-source.rst, docs/source/lib/install\_datatypes.rst, docs/html/ux-research-design/contribute.md\\
        `{}`{}`\\
        A reasonable output is:\\
        `{}`{}`\\
        <ANSWER>: readme.md, contributing.md, docs/maintaining/installing/install-from-source.rst,\\
        <REASONING>: The files readme.md and contributing.md commonly contain installation and testing instructions, while docs/maintaining/installing/install-from-source.rst is likely related to installation as the name suggests\\
        `{}`{}`\\
        Here are the file names\\
        `{}`{}`\\
        <file 1>, <file 2>, \ldots, <file k>\\
        `{}`{}`\\
        Please read the names carefully, ask yourself the purpose of each file based on the name before including it in your response. Use the given format for your answer and please do not add any extra comment or text.}
    \vspace{-4mm}
    \caption{Prompt for choosing relevant files to installation and testing}
    \vspace{-4mm}
    \label{fig:prompt_files_suggestion}
\end{figure}

\begin{figure}[H]
    \centering
    \prompt{Prompt to suggest external sources of information}{
        You are a senior developer contributing to the GitHub project at www.github.com/<repo\_id> by solving issues. Your goal is to install the repository in development mode and run its tests.\\
You have created a Docker environment with Ubuntu, and now you are searching for the installation instructions and test commands.\\ 
I will provide you with the content of common repository files (e.g., README.md, CONTRIBUTING.md). Your task is to analyze the provided text and identify all external links that contain relevant information to\\
1. Installation instructions for this project.\\
2. Test commands or instructions for running the tests for this project.\\
3. Contribution guidelines.\\

Please provide the links you found following the criteria below.\\
a. Exclude links to general\-purpose documentation for external tools (e.g., Tox, Pytest, or other frameworks/libraries).\\
b. If you are unsure about the relevance of a link, better include it.\\
c. Order the links from most to least relevant.\\
d. Do not add any comment or text.\\
e. Use the following format:\\
LINK: <LINK 1>\\
LINK: <LINK 2>\\
\ldots
LINK: <LINK n>\\
Here is the text:\\
'{}'{}'\\
<text\_content>\\
'{}'{}'}
    \vspace{-4mm}
    \caption{Prompt to suggest potentially relevant external sources}
    \vspace{-4mm}
    \label{fig:prompt_for_external_sourcers}
\end{figure}

\begin{figure}[H]
    \centering
    \prompt{Prompt to determine importance of a url content}{You are a senior developer working on the GitHub project at www.github.com/<repo\_id>. You have set up a Docker environment with Ubuntu, and now your goal is
     to install the repository in development mode and run its tests.\\
    Your task is to carefully review the content of the following link: <current\_link>, and determine if it includes installation instructions or test commands for the <repo\_id> project.\\
    Please follow these steps:\\
    1. Look carefully in the provided content for any potential installation commands or test commands related to the <repo\_id> project.\\
    2. Ask yourself if the located instructions are reasonable, legitimate and can be practically executed to install or to test the <repo\_id> project only.\\
    Please provide your answer using the following format:\\
    INSTALLATION/TEST COMMANDS: <TRUE|FALSE>\\
    REASONING: <REASONING>\\
    **Important Notes**\\
    \- Answer with TRUE only if the content explicitly includes valid and usable installation or test commands.\\
    \- If you do not find any relevant commands, or if the instructions are vague, ambiguous, impractical, or unrelated answer FALSE.\\
    \- When in doubt, answer FALSE.\\
    Content of the link <current\_link>:\\
    '{}'{}'\\
    <clean\_content>
    '{}'{}'}
    \vspace{-4mm}
    \caption{Prompt for determining if a link is relevant to installation and testing in the extraction phase of the \agent}
    \vspace{-4mm}
    \label{fig:prompt_for_importance}
\end{figure}
\begin{figure}[H]
    \centering
    \prompt{Extract Install Command Prompt}{
You are a senior developer working on the project located at www.github.com/<repo\_id>. You have created a Docker environment with Ubuntu, cloned the repository, and navigated to the directory <repo\_dir>.

Your next step is to install the project in development mode, which is intended for active development and testing. I'll provide you with important text files (e.g., README.md) and important continuous integration (CI) configuration files, which typically contain instructions for developers on installation and testing. The format provided will be the file name followed by its content.

Your task is to identify and return the bash commands necessary for the correct installation of the repository. This includes system dependencies, project installation in development mode, and any prerequisites or configuration commands.\\
\\
** IMPORTANT NOTES **\\
1. Include system dependencies installation commands required for the project (e.g., via apt, yum, curl, etc.).\\
2. Include installation commands necessary for setting up the project in development mode.\\
3. Include prerequisites installation and configuration commands, such as those for npm or any other required setup.\\
3. If comprehensive installation instructions are provided, return them without any modifications.\\
4. Only exclude commands related to creating or activating virtual environments.\\

The returned commands should meet the following criteria:\\
1. Enclosed in quotes.\\
2. Focused strictly on commands necessary for both system dependency installation and development-mode installation of the project.\\
3. Free from any comments or text.\\
4. Accurate and executable without errors.\\

If no installation commands are present, return NONE.\\
Here is the text:\\
`{}`{}`\\
<context>\\
`{}`{}`\\
Take your time to carefully analyze the content. Make sure that your response includes only the necessary installation bash commands. Ask yourself if the provided content is sufficient for installation. And for each command, ask yourself what's the purpose of the command and if it is necessary.\\
An example of the expected response is:\\
`{}`{}`bash\\
install\_command\_1\\
install\_command\_2\\
`{}`{}`\\
Please provide the installation commands in the above specified format.}
    \vspace{-4mm}
    \caption{Prompt used for extraction of installation commands in extraction phase of \agent}
    \vspace{-4mm}
    \label{fig:prompt_installation_extraction}
    \end{figure}

    \begin{figure}[H]
        \centering
        \prompt{Extract Test Command Prompt}{
            You are a senior developer working on the www.github.com/<repo\_id> project. You have created a Docker environment with Ubuntu, cloned the repository, and installed it in development mode (meant for active development and testing).\\
            You are now inside the <repo\_dir> directory and your next goal is to run the unit tests. I will provide you with some important text files (e.g., README.md) and important continuous integration (CI) congiguration files, which typically include instructions for running tests. The format provided will be the file name followed by its content.\\
            Your task is to identify and return the exact bash commands required to run the tests.\\
            The returned commands should meet the following criteria:\\
            1. Enclosed in quotes.\\
            2. Free from any comments or text.\\
            3. Accurate and executable without errors.\\
            If no test commands are present, return NONE.\\
            Here is the text:\\
            `{}`{}`\\
            <context>\\
            `{}`{}`\\
            Take your time to analyze the content carefully. Ensure that only the necessary bash commands for running the tests are included. Ask yourself the purpose of each command before including it in your response.\\
            An example of the expected response is:\\
            `{}`{}`bash\\
            test\_command\_1\\
            test\_command\_2\\
            `{}`{}`\\
            Please provide the test commands in the above specified format.}
        \vspace{-4mm}
        \caption{Prompt used for extraction of test commands in the extraction phase of \agent}
        \vspace{-4mm}
        \label{fig:prompt_test_extraction}
        \end{figure}

        \begin{figure}[H]
            \centering
            \prompt{Prompt for determining error causes}{You are a developer working on the project at www.github.com/<repo\_id>. You created an environment with python version <python\_version>. Your goal is to install the repository in development mode (meant for active development and testing) and run the unit tests.\\
            The installation commands are:\\
            `{}`{}`bash\\
            <install\_command\_1>\\
            <install\_command\_2>\\
            \ldots\\
            <install\_command\_k>\\
            `{}`{}`\\
            The testing commands are:\\
            `{}`{}`bash\\
            <test\_command\_1>\\
            <test\_command\_2>\\
            \ldots\\
            <test\_command\_k>\\
            `{}`{}`\\
            You received the following error message after executing the command <error\_command>:\\
            '{}'{}'\\
            <error\_message>\\
            '{}'{}'\\
            Your task is to analyze the error message and determine its causes.\\
            You can return one of the following answers:\\
            1. <PYTHON>, if the error is caused by incompatibilities between the python version and any used package.\\
            2. <INSTALLATION>, if the error is caused by an installation command or is related to any missing package, regardless if it a testing related framework or not. All the required packages must be installed in the installation phase.\\
            3. <TESTING>, if the error is caused by any testing command (e.g., an invalid flag in the test command)\\
            4. <UNDECIDABLE>, if you cannot determine what causes the error.\\
            Please read the error message carefully and try to spot the commands that are responsible for the error. Always provide the reasoning for your answer.\\
            Use the following format:\\
            RESULT: <PYTHON, INSTALLATION, TESTING, UNDECIDABLE>\\
            REASONING: <YOUR REASONING>\\
            }
            \vspace{-4mm}
            \caption{Prompt for determining the error cause in the iterative improvement phase of the \agent}
            \vspace{-4mm}
            \label{fig:prompt_error_cause}
            \end{figure}
    
        \begin{figure}[H]
            \centering
            \prompt{Prompt for fixing python version}{You are a senior developer working on the project at www.github.com/<repo\_id>. Your goal is to install the repository in development mode (meant for active development and testing) and run the unit tests.\\
            You created an environment with python version <python\_version>, but you are unsure if the python version is correct.\\
            You received the following error message while testing the repository:\\
            '{}'{}'\\
            <error\_message>\\
            '{}'{}'\\
            A senior software developer colleague has provided an explanation of why things are not working as expected with the current commands:\\
            <Reasoning from the answer to the prompt for determining the error cause>. Use his reasoning to resolve the current error we are facing.\\
            Your task is to determine a compatible Python version for the current state of the repository. Carefully read the error message and identify the most suitable Python version.\\
            Please follow this answer format:\\
            1. Return <NONE> if the error is unrelated to the Python version or you cannot determine a compatible version.\\
            2. If a specific Python version is compatible, return only the version number (e.g., \"2.7\").\\
            3. Do not include any additional comments or text in your response.\\
            }
            \vspace{-4mm}
            \caption{Prompt for fixing python version used in the iterative improvement phase of \agent}
            \vspace{-4mm}
            \label{fig:prompt_for_python_version}
            \end{figure}
    \begin{figure}[H]
        \centering
        \prompt{Prompt for fixing installation commands 1}{You are a senior developer working on the project at www.github.com/<repo\_id>. You are working in an enviroment with python version <python\_version>. You have attempted to install the repository in development mode (meant for active development and testing) using the following bash commands:\\
        `{}`{}`bash\\
        <install\_command\_1>\\
        <install\_command\_2>\\
        \ldots\\
        <install\_command\_n>\\
        `{}`{}`\\
        However, the command <error\_command> failed and we received the following error message:\\
        '{}'{}'\\
        <error\_message>
        '{}'{}'\\
        Your task is to fix the above error. Think carefully what causes the error and try to spot the commands that are responsible for it. Please provide the updated installation steps in a bash code block, following these rules:\\
        1. You have to use always uv pip instead of regular pip.\\
        2. Return <NONE> if you can not fix the command.\\
        3. Do not add any comments or text.\\
        For example:\\
        `{}`{}`bash\\
        apt-get install -y <package\_name>\\
        uv pip install -r requirements.txt\\
        `{}`{}`}
        \vspace{-4mm}
        \caption{Prompt for fixing the installation commands used in the iterative improvement phase of \agent when the error occurs in the building process of containers}
        \vspace{-4mm}
        \label{fig:prompt_installation_fixing}
        \end{figure}

    \begin{figure}[H]
        \centering
        \prompt{Prompt for fixing installation commands 2}{You are a senior developer working on the project at www.github.com/<repo\_id>. You tried to install the repository in development mode, which is intended for active development and testing, however the installation failed.\\
        You are working in an enviroment with python version <python> and you tried to use the following bash commands for the installation:\\
        `{}`{}`bash\\
        <install\_command\_1>\\
        <install\_command\_2>\\
        \ldots\\
        <install\_command\_n>\\
        `{}`{}`\\
        During the execution of these commands, you received the following error message:
        '{}'{}'\\
        <error\_message>
        '{}'{}'\\
        A senior software developer colleague has provided an explanation of why things are not working as expected with the current commands:\\
        <Reasoning from the answer to the prompt for determining the error cause>. Use his reasoning to resolve the current error we are facing.\\
        Your task is to carefully read the error message and determine which commands are causing the error. Reason about every command if it is causing the error. If you conclude that the problem is related to any of the commands, update the installation bash script to solve the problem. Note that you can also add new commands to fix the problem.
        If you decide to update the installation bash script you have to follow these rules:\\
        1. Provide the updated installation steps in a bash code block.\\
        2. Use uv pip instead of regular pip.\\
        2. Return NONE if the error is not related to the installation steps or you are not able to fix it.\\
        3. Do not add any comments or text.\\
        For example:\\ 
        The initial installation command is:\\
        `{}`{}`bash\\
        uv pip install \.\\
        `{}`{}`\\
        However, the error message states that the <package\_name> package is not installed. Then you would update the installation command to:\\
        `{}`{}`bash\\
        uv pip install \.\\
        uv pip install <package\_name>
        `{}`{}`}
        \vspace{-4mm}
        \caption{Prompt for fixing the installation commands used in the iterative improvement phase of \agent}
        \vspace{-4mm}
        \label{fig:prompt_installation_fixing_v2}
        \end{figure}
    
    \begin{figure}[H]
        \centering
        \prompt{Prompt for fixing testing commands}{You are a senior developer working on the project at www.github.com/<repo\_id>. You installed the repository in an enviroment with python version <python\_version> and now you are trying to run the unit tests.\\
        You run the tests using the following bash commands:\\
        `{}`{}`bash\\
        <test\_command\_1>\\
        <test\_command\_2>\\
        \ldots\\
        <test\_command\_k>\\
        `{}`{}`\\
        However, at the moment we receive the following error message:\\
        '{}'{}'
        <error\_message>
        '{}'{}'\\
        A senior software developer colleague has provided an explanation of why things are not working as expected with the current commands:\\
        <Reasoning from the answer to the prompt for determining the error cause>. Use his reasoning to resolve the current error we are facing.\\
        Your task is to read the produced error message carefully, determine what the problem is and try to fix it. Ask yourself which test command could cause this problem. If you conclude that the problem is related to the test commands, update the test commands to solve the problem.\\
        Please provide the updated test commnds in a bash code block, following these rules:\\
        1. You have to always use uv pip instead of regular pip.\\
        2. Return NONE if the error is not related to the test command or you cannot fix it.\\
        3. Do not add any comments or text.\\
        4. Add a command only if you are sure that it is correct.\\
        For example: The initial testing command was:\\
        `{}`{}`bash\\
        pytest test\_file.py \-\-run \-all
        `{}`{}`
        However, if in this case we would need the flag '-v' and the maximal number of failing tests to be 1, we would have to correct the command to:\\
        `{}`{}`bash\\
        pytest test\_file.py \-\-maxfail=1 \-v
        `{}`{}`}
        \vspace{-4mm}
        \caption{Prompt for fixing the installation commands used in the iterative improvement phase of \agent}
        \vspace{-4mm}
        \label{fig:prompt_installation_fixing_v2}
        \end{figure}

\newpage
\section{Appendix -- Dataset Details} \label{app:repo_list}

Below, we list all repositories along with the number of corresponding tasks in \swab.

\prompt{\swab -- Repositories}{
    \begin{multicols}{2}
        \begin{enumerate}
            \scriptsize
            \item \code{iterative/dvc} -- 42
            \item \code{streamlink/streamlink} -- 35
            \item \code{spack/spack} -- 35
            \item \code{PrefectHQ/prefect} -- 34
            \item \code{xonsh/xonsh} -- 32
            \item \code{mitmproxy/mitmproxy} -- 31
            \item \code{python-pillow/Pillow} -- 29
            \item \code{mkdocs/mkdocs} -- 23
            \item \code{hynek/structlog} -- 22
            \item \code{pallets/click} -- 21
            \item \code{locustio/locust} -- 20
            \item \code{jpadilla/pyjwt} -- 17
            \item \code{elastic/elasticsearch-dsl-py} -- 17
            \item \code{pallets-eco/wtforms} -- 17
            \item \code{ipython/ipython} -- 16
            \item \code{python-poetry/poetry} -- 15
            \item \code{conan-io/conan} -- 15
            \item \code{sabnzbd/sabnzbd} -- 14
            \item \code{Zulko/moviepy} -- 14
            \item \code{nvbn/thefuck} -- 12
            \item \code{arrow-py/arrow} -- 11
            \item \code{benoitc/gunicorn} -- 8
            \item \code{cookiecutter/cookiecutter} -- 8
            \item \code{pypa/pipenv} -- 7
            \item \code{graphql-python/graphene} -- 6
            \item \code{pypa/bandersnatch} -- 5
            \item \code{AtsushiSakai/PythonRobotics} -- 4
            \item \code{hynek/doc2dash} -- 3
            \item \code{PythonCharmers/python-future} -- 3
            \item \code{aimhubio/aim} -- 2
            \item \code{dbcli/pgcli} -- 2
            \item \code{geopython/pycsw} -- 2
            \item \code{dbader/schedule} -- 2
            \item \code{kibitzr/kibitzr} -- 1
            \item \code{getnikola/nikola} -- 1
            \item \code{geopy/geopy} -- 1
            \item \code{Maratyszcza/PeachPy} -- 1
            \item \code{gawel/pyquery} -- 1
            \item \code{Suor/funcy} -- 1
            \item \code{simonw/datasette} -- 1
            \item \code{cowrie/cowrie} -- 1
            \item \code{pypa/pip} -- 1
            \item \code{StevenBlack/hosts} -- 1
            \item \code{jupyter/nbgrader} -- 1
        \end{enumerate}
    \end{multicols}
}

Below, we list all repositories along with the number of corresponding tasks in \sweeb.
\prompt{\sweeb -- Repositories Part I}{
    \begin{multicols}{2}
        \begin{enumerate}
            \scriptsize
            \item \code{python-attrs/attrs} -- 9
            \item \code{dgasmith/opt\_einsum} -- 9
            \item \code{jazzband/tablib} -- 8
            \item \code{MartinThoma/flake8-simplify} -- 8
            \item \code{matthewwithanm/python-markdownify} -- 8
            \item \code{stephenhillier/starlette\_exporter} -- 8
            \item \code{sciunto-org/python-bibtexparser} -- 8
            \item \code{davidhalter/parso} -- 8
            \item \code{marshmallow-code/flask-smorest} -- 7
            \item \code{adamchainz/blacken-docs} -- 7
            \item \code{MarketSquare/robotframework-tidy} -- 7
            \item \code{lundberg/respx} -- 7
            \item \code{seperman/deepdiff} -- 7
            \item \code{Stranger6667/hypothesis-graphql} -- 7
            \item \code{cantools/cantools} -- 7
            \item \code{didix21/mdutils} -- 7
            \item \code{marshmallow-code/apispec} -- 7
            \item \code{softlayer/softlayer-python} -- 6
            \item \code{gorakhargosh/watchdog} -- 6
            \item \code{pygments/pygments} -- 6
            \item \code{dask-contrib/dask-histogram} -- 6
            \item \code{andialbrecht/sqlparse} -- 6
            \item \code{mirumee/ariadne} -- 6
            \item \code{tableau/tabcmd} -- 6
            \item \code{gerrymanoim/exchange\_calendars} -- 5
            \item \code{snowplow/snowplow-python-tracker} -- 5
            \item \code{joerick/pyinstrument} -- 5
            \item \code{scikit-rf/scikit-rf} -- 5
            \item \code{matthewwardrop/formulaic} -- 5
            \item \code{laspy/laspy} -- 5
            \item \code{python-control/python-control} -- 5
            \item \code{mwouts/itables} -- 5
            \item \code{AzureAD/microsoft-authentication-library-for-python} -- 5
            \item \code{firebase/firebase-admin-python} -- 5
            \item \code{ethereum/eth-account} -- 5
            \item \code{davidhalter/jedi} -- 5
            \item \code{agronholm/typeguard} -- 5
            \item \code{Delgan/loguru} -- 5
            \item \code{pytransitions/transitions} -- 5
            \item \code{lovasoa/marshmallow\_dataclass} -- 5
            \item \code{aio-libs/yarl} -- 5
            \item \code{PyCQA/pyflakes} -- 5
            \item \code{python/importlib\_metadata} -- 5
            \item \code{konradhalas/dacite} -- 5
            \item \code{ilevkivskyi/typing\_inspect} -- 5
            \item \code{jupyter/jupyter\_core} -- 5
            \item \code{getsentry/responses} -- 5
            \item \code{beartype/plum} -- 4
            \item \code{open2c/bioframe} -- 4
            \item \code{developmentseed/morecantile} -- 4
            \item \code{nats-io/nats.py} -- 4
            \item \code{nipy/nipype} -- 4
            \item \code{python-quantities/python-quantities} -- 4
            \item \code{stac-utils/pystac-client} -- 4
            \item \code{luolingchun/flask-openapi3} -- 4
            \item \code{sayanarijit/expandvars} -- 4
            \item \code{jpadilla/pyjwt} -- 4
            \item \code{NowanIlfideme/pydantic-yaml} -- 4
            \item \code{john-kurkowski/tldextract} -- 4
            \item \code{geopandas/geopandas} -- 4
            \item \code{cloudevents/sdk-python} -- 4
            \item \code{jupyter/nbformat} -- 4
            \item \code{matthew-brett/delocate} -- 4
            \item \code{iterative/shtab} -- 4
            \item \code{jsonpickle/jsonpickle} -- 4
            \item \code{ethereum/eth-utils} -- 4
            \item \code{mhe/pynrrd} -- 4
            \item \code{adamjstewart/fiscalyear} -- 4
            \item \code{pytest-dev/pytest-xdist} -- 4
            \item \code{facelessuser/wcmatch} -- 4
            \item \code{scikit-hep/awkward} -- 4
            \item \code{tomplus/kubernetes\_asyncio} -- 4
            \item \code{ipython/traitlets} -- 4
            \item \code{David-Wobrock/sqlvalidator} -- 4
            \item \code{omry/omegaconf} -- 4
            \item \code{python-lsp/python-lsp-server} -- 4
            \item \code{cogeotiff/rio-tiler} -- 3
            \item \code{wjohnson/pyapacheatlas} -- 3
        \end{enumerate}
    \end{multicols}
}
\prompt{\sweeb~-- Repositories Part II}{
    \begin{multicols}{2}
        \begin{enumerate}
            \setcounter{enumi}{78}
            \scriptsize
            \item \code{adamchainz/django-htmx} -- 3
            \item \code{mwclient/mwclient} -- 3
            \item \code{executablebooks/sphinx-book-theme} -- 3
            \item \code{scikit-hep/vector} -- 3
            \item \code{patrick-kidger/equinox} -- 3
            \item \code{christiansandberg/canopen} -- 3
            \item \code{regebro/pyroma} -- 3
            \item \code{nephila/giturlparse} -- 3
            \item \code{cookiecutter/cookiecutter} -- 3
            \item \code{serge-sans-paille/pythran} -- 3
            \item \code{tomasvotava/fastapi-sso} -- 3
            \item \code{jsvine/pdfplumber} -- 3
            \item \code{scrapy/protego} -- 3
            \item \code{SmileyChris/django-countries} -- 3
            \item \code{cscorley/whatthepatch} -- 3
            \item \code{pythological/kanren} -- 3
            \item \code{pypa/virtualenv} -- 3
            \item \code{fastavro/fastavro} -- 3
            \item \code{marshmallow-code/marshmallow-sqlalchemy} -- 3
            \item \code{gazpachoking/jsonref} -- 3
            \item \code{lepture/mistune} -- 3
            \item \code{scikit-learn-contrib/category\_encoders} -- 3
            \item \code{simonw/sqlite-utils} -- 3
            \item \code{executablebooks/mdit-py-plugins} -- 3
            \item \code{tsutsu3/linkify-it-py} -- 3
            \item \code{hhatto/autopep8} -- 3
            \item \code{cubewise-code/mdxpy} -- 3
            \item \code{joblib/joblib} -- 3
            \item \code{python-trio/trio-typing} -- 3
            \item \code{nalepae/pandarallel} -- 3
            \item \code{tableau/server-client-python} -- 3
            \item \code{r1chardj0n3s/parse} -- 3
            \item \code{ipython/ipython} -- 3
            \item \code{pypa/readme\_renderer} -- 3
            \item \code{jaraco/zipp} -- 3
            \item \code{docker/docker-py} -- 3
            \item \code{joshy/striprtf} -- 3
            \item \code{googleapis/python-pubsub} -- 3
            \item \code{TylerYep/torchinfo} -- 3
            \item \code{scrapy/w3lib} -- 3
            \item \code{googleapis/google-auth-library-python-oauthlib} -- 3
            \item \code{agronholm/cbor2} -- 3
            \item \code{weiwei/junitparser} -- 3
            \item \code{conan-io/conan} -- 3
            \item \code{python/importlib\_resources} -- 3
            \item \code{timvink/mkdocs-git-authors-plugin} -- 3
            \item \code{agronholm/exceptiongroup} -- 3
            \item \code{magmax/python-inquirer} -- 3
            \item \code{PrefectHQ/prefect} -- 3
            \item \code{Yelp/detect-secrets} -- 3
            \item \code{Chilipp/autodocsumm} -- 3
            \item \code{jaraco/keyring} -- 3
            \item \code{Pylons/waitress} -- 3
            \item \code{pypa/setuptools} -- 3
            \item \code{barrust/pyspellchecker} -- 2
            \item \code{bluesky/ophyd} -- 2
            \item \code{OpenMath/py-openmath} -- 2
            \item \code{readthedocs/sphinx-notfound-page} -- 2
            \item \code{canonical/operator} -- 2
            \item \code{ekzhu/datasketch} -- 2
            \item \code{dhatim/python-license-check} -- 2
            \item \code{Shoobx/xmldiff} -- 2
            \item \code{ewels/rich-click} -- 2
            \item \code{jaraco/path} -- 2
            \item \code{yu-iskw/dbt-artifacts-parser} -- 2
            \item \code{symerio/pgeocode} -- 2
            \item \code{daggaz/json-stream} -- 2
            \item \code{jazzband/dj-database-url} -- 2
            \item \code{nipunsadvilkar/pySBD} -- 2
            \item \code{adamchainz/django-linear-migrations} -- 2
            \item \code{mwouts/jupytext} -- 2
            \item \code{MrBin99/django-vite} -- 2
            \item \code{ml31415/numpy-groupies} -- 2
            \item \code{regebro/svg.path} -- 2
            \item \code{gmr/flatdict} -- 2
            \item \code{aws-samples/sample-python-helper-aws-appconfig} -- 2
            \item \code{behave/behave} -- 2
        \end{enumerate}
    \end{multicols}
}
\prompt{\sweeb~-- Repositories Part III}{
    \begin{multicols}{2}
        \begin{enumerate}
            \setcounter{enumi}{155}
            \scriptsize
            \item \code{thesimj/envyaml} -- 2
            \item \code{codingjoe/django-select2} -- 2
            \item \code{allisson/python-simple-rest-client} -- 2
            \item \code{christianhelle/autofaker} -- 2
            \item \code{esphome/aioesphomeapi} -- 2
            \item \code{oauthlib/oauthlib} -- 2
            \item \code{rustedpy/result} -- 2
            \item \code{graphql-python/graphene} -- 2
            \item \code{benmoran56/esper} -- 2
            \item \code{eerimoq/bincopy} -- 2
            \item \code{keleshev/schema} -- 2
            \item \code{PyCQA/flake8} -- 2
            \item \code{kjd/idna} -- 2
            \item \code{jupyter/nbconvert} -- 2
            \item \code{scikit-hep/hist} -- 2
            \item \code{spulec/freezegun} -- 2
            \item \code{jupyter/nbclient} -- 2
            \item \code{PythonCharmers/python-future} -- 2
            \item \code{tortoise/pypika-tortoise} -- 2
            \item \code{rthalley/dnspython} -- 2
            \item \code{mkaranasou/pyaml\_env} -- 2
            \item \code{terraform-compliance/cli} -- 2
            \item \code{googleapis/python-firestore} -- 2
            \item \code{googleapis/python-api-core} -- 2
            \item \code{scrapy/cssselect} -- 2
            \item \code{python-humanize/humanize} -- 2
            \item \code{jdepoix/youtube-transcript-api} -- 2
            \item \code{dedupeio/dedupe} -- 2
            \item \code{databricks/databricks-cli} -- 2
            \item \code{bluesky/event-model} -- 2
            \item \code{workos/workos-python} -- 2
            \item \code{kynan/nbstripout} -- 2
            \item \code{assertpy/assertpy} -- 2
            \item \code{dbt-labs/hologram} -- 2
            \item \code{sendgrid/python-http-client} -- 2
            \item \code{keis/base58} -- 2
            \item \code{attwad/python-osc} -- 2
            \item \code{wireservice/csvkit} -- 2
            \item \code{adamchainz/time-machine} -- 2
            \item \code{MagicStack/immutables} -- 2
            \item \code{vinitkumar/json2xml} -- 2
            \item \code{frispete/keyrings.cryptfile} -- 2
            \item \code{swansonk14/typed-argument-parser} -- 2
            \item \code{scottwernervt/favicon} -- 2
            \item \code{slackapi/python-slack-sdk} -- 2
            \item \code{nginxinc/crossplane} -- 2
            \item \code{hetznercloud/hcloud-python} -- 2
            \item \code{dbader/schedule} -- 2
            \item \code{amplify-education/python-hcl2} -- 2
            \item \code{jazzband/contextlib2} -- 2
            \item \code{theskumar/python-dotenv} -- 2
            \item \code{raimon49/pip-licenses} -- 2
            \item \code{locustio/locust} -- 2
            \item \code{astanin/python-tabulate} -- 2
            \item \code{alecthomas/voluptuous} -- 2
            \item \code{django-crispy-forms/crispy-bootstrap5} -- 2
            \item \code{geospace-code/pymap3d} -- 2
            \item \code{tedder/requests-aws4auth} -- 2
            \item \code{pyvisa/pyvisa-py} -- 1
            \item \code{nithinmurali/pygsheets} -- 1
            \item \code{mlenzen/collections-extended} -- 1
            \item \code{emcconville/wand} -- 1
            \item \code{rsalmei/alive-progress} -- 1
            \item \code{rycus86/prometheus\_flask\_exporter} -- 1
            \item \code{fastapi-users/fastapi-users} -- 1
            \item \code{google/mobly} -- 1
            \item \code{scrapy/itemadapter} -- 1
            \item \code{ncclient/ncclient} -- 1
            \item \code{google/duet} -- 1
            \item \code{di/calver} -- 1
            \item \code{beancount/smart\_importer} -- 1
            \item \code{bridgecrewio/python-hcl2} -- 1
            \item \code{construct/construct} -- 1
            \item \code{devrimcavusoglu/pybboxes} -- 1
            \item \code{richardpenman/whois} -- 1
            \item \code{cvxpy/cvxpy} -- 1
            \item \code{elastic/ecs-logging-python} -- 1
            \item \code{pythonarcade/pytiled\_parser} -- 1
            \item \code{astropy/extension-helpers} -- 1
        \end{enumerate}
    \end{multicols}
}
\prompt{\sweeb~-- Repositories Part IV}{
    \begin{multicols}{2}
        \begin{enumerate}
            \setcounter{enumi}{234}
            \scriptsize
            \item \code{SAP/python-pyodata} -- 1
            \item \code{Azure/azure-functions-durable-python} -- 1
            \item \code{IdentityPython/djangosaml2} -- 1
            \item \code{jwodder/check-wheel-contents} -- 1
            \item \code{Zulko/moviepy} -- 1
            \item \code{xhtml2pdf/xhtml2pdf} -- 1
            \item \code{cknd/stackprinter} -- 1
            \item \code{guillp/jwskate} -- 1
            \item \code{jmcarp/flask-apispec} -- 1
            \item \code{timofurrer/colorful} -- 1
            \item \code{miso-belica/sumy} -- 1
            \item \code{kvesteri/intervals} -- 1
            \item \code{marcotcr/lime} -- 1
            \item \code{wkentaro/gdown} -- 1
            \item \code{realpython/codetiming} -- 1
            \item \code{jaraco/tempora} -- 1
            \item \code{jendrikseipp/vulture} -- 1
            \item \code{pycontribs/ruyaml} -- 1
            \item \code{albumentations-team/albumentations} -- 1
            \item \code{nose-devs/nose2} -- 1
            \item \code{jongracecox/anybadge} -- 1
            \item \code{patrys/httmock} -- 1
            \item \code{maxfischer2781/asyncstdlib} -- 1
            \item \code{pgzip/pgzip} -- 1
            \item \code{arvkevi/kneed} -- 1
            \item \code{rasterio/affine} -- 1
            \item \code{circus-tent/circus} -- 1
            \item \code{xchwarze/samsung-tv-ws-api} -- 1
            \item \code{jaraco/portend} -- 1
            \item \code{fabiocaccamo/python-benedict} -- 1
            \item \code{numpy/numpy-financial} -- 1
            \item \code{praw-dev/prawcore} -- 1
            \item \code{scipy/oldest-supported-numpy} -- 1
            \item \code{logtail/logtail-python} -- 1
            \item \code{polkascan/py-scale-codec} -- 1
            \item \code{Knio/pynmea2} -- 1
            \item \code{jazzband/django-configurations} -- 1
            \item \code{allenai/cached\_path} -- 1
            \item \code{click-contrib/click-aliases} -- 1
            \item \code{Pylons/hupper} -- 1
            \item \code{cloudscale-ch/cloudscale-python-sdk} -- 1
            \item \code{alessandromaggio/pythonping} -- 1
            \item \code{imageio/imageio-ffmpeg} -- 1
            \item \code{podhmo/python-node-semver} -- 1
            \item \code{netbox-community/pynetbox} -- 1
            \item \code{kumar303/mohawk} -- 1
            \item \code{SpamScope/mail-parser} -- 1
            \item \code{perrygeo/python-rasterstats} -- 1
            \item \code{pahaz/sshtunnel} -- 1
            \item \code{python-hyper/h11} -- 1
            \item \code{razorpay/razorpay-python} -- 1
            \item \code{zeroSteiner/rule-engine} -- 1
            \item \code{mocobeta/janome} -- 1
            \item \code{glut23/webvtt-py} -- 1
            \item \code{benoitc/gunicorn} -- 1
            \item \code{mcmtroffaes/pybtex-docutils} -- 1
            \item \code{alexmojaki/executing} -- 1
            \item \code{sigmavirus24/github3.py} -- 1
            \item \code{ccpem/mrcfile} -- 1
            \item \code{csinva/imodels} -- 1
            \item \code{click-contrib/click-help-colors} -- 1
            \item \code{srossross/rpmfile} -- 1
            \item \code{hgrecco/pint} -- 1
            \item \code{django-ses/django-ses} -- 1
            \item \code{gmr/pamqp} -- 1
            \item \code{spotify/annoy} -- 1
            \item \code{PyCQA/pycodestyle} -- 1
            \item \code{regebro/tzlocal} -- 1
            \item \code{mapado/haversine} -- 1
            \item \code{scientific-python/lazy-loader} -- 1
            \item \code{grappa-py/grappa} -- 1
            \item \code{flexmock/flexmock} -- 1
            \item \code{jg-rp/liquid} -- 1
            \item \code{prompt-toolkit/python-prompt-toolkit} -- 1
            \item \code{jaraco/jaraco.context} -- 1
            \item \code{aio-libs/multidict} -- 1
            \item \code{rsheftel/pandas\_market\_calendars} -- 1
        \end{enumerate}
    \end{multicols}
}
\prompt{\sweeb~-- Repositories Part V}{
    \begin{multicols}{2}
        \begin{enumerate}
            \setcounter{enumi}{311}
            \scriptsize
            \item \code{mkdocs/mkdocs} -- 1
            \item \code{websocket-client/websocket-client} -- 1
            \item \code{DataDog/datadog-lambda-python} -- 1
            \item \code{iterative/dvclive} -- 1
            \item \code{cogeotiff/rio-cogeo} -- 1
            \item \code{erikrose/parsimonious} -- 1
            \item \code{facelessuser/pymdown-extensions} -- 1
            \item \code{pypa/build} -- 1
            \item \code{mkdocs/mkdocs-redirects} -- 1
            \item \code{dlint-py/dlint} -- 1
            \item \code{klen/peewee\_migrate} -- 1
            \item \code{afq984/python-cxxfilt} -- 1
            \item \code{kinverarity1/lasio} -- 1
            \item \code{Turbo87/utm} -- 1
            \item \code{django/daphne} -- 1
            \item \code{executablebooks/sphinx-design} -- 1
            \item \code{interpretml/slicer} -- 1
            \item \code{google/yapf} -- 1
            \item \code{sensein/etelemetry-client} -- 1
            \item \code{MKuranowski/aiocsv} -- 1
            \item \code{executablebooks/sphinx-tabs} -- 1
            \item \code{pexpect/pexpect} -- 1
            \item \code{pythological/etuples} -- 1
            \item \code{frankie567/httpx-oauth} -- 1
            \item \code{sarugaku/resolvelib} -- 1
            \item \code{python273/telegraph} -- 1
            \item \code{boolangery/py-lua-parser} -- 1
            \item \code{Electrostatics/mmcif\_pdbx} -- 1
            \item \code{pyca/service-identity} -- 1
            \item \code{diff-match-patch-python/diff-match-patch} -- 1
            \item \code{xlwings/jsondiff} -- 1
            \item \code{mapbox/cligj} -- 1
            \item \code{cthoyt/pystow} -- 1
            \item \code{Rapptz/discord.py} -- 1
            \item \code{gahjelle/pyplugs} -- 1
            \item \code{Colin-b/pytest\_httpx} -- 1
            \item \code{LLNL/certipy} -- 1
            \item \code{spec-first/connexion} -- 1
            \item \code{Yelp/bravado} -- 1
            \item \code{mkorpela/pabot} -- 1
            \item \code{scrapy/parsel} -- 1
            \item \code{alexmojaki/pure\_eval} -- 1
            \item \code{graphql-python/graphql-core} -- 1
            \item \code{joke2k/faker} -- 1
            \item \code{averbis/averbis-python-api} -- 1
            \item \code{jupyter/jupyter\_client} -- 1
            \item \code{jaraco/inflect} -- 1
            \item \code{GreyZmeem/python-logging-loki} -- 1
            \item \code{suminb/base62} -- 1
            \item \code{youknowone/wirerope} -- 1
            \item \code{xnuinside/simple-ddl-parser} -- 1
            \item \code{executablebooks/sphinx-thebe} -- 1
            \item \code{Pylons/webob} -- 1
            \item \code{SethMMorton/fastnumbers} -- 1
            \item \code{python-semver/python-semver} -- 1
        \end{enumerate}
    \end{multicols}
}

\fi

\end{document}